\documentclass[prd,preprint,showpacs,eqsecnum,floatfix]{revtex4}
\usepackage{amsfonts}
\usepackage{amssymb}
\usepackage{graphicx}
\usepackage{pstricks}

\newcommand{\BOX}{\hbox {$\sqcap$ \kern -1em $\sqcup$}}

\newcommand{\be}{\begin{equation}}
\newcommand{\ee}{\end{equation}}
\newcommand{\ba}{\begin{eqnarray}}
\newcommand{\ea}{\end{eqnarray}}
\newcommand{\ban}{\begin{eqnarray*}}
\newcommand{\bea}{\begin{eqnarray}}
\newcommand{\eea}{\end{eqnarray}}
\newcommand{\ean}{\end{eqnarray*}}
\newcommand{\barr}{\begin{array}}
\newcommand{\earr}{\end{array}}
\newcommand{\bra}[1]{\langle #1|}
\newcommand{\ket}[1]{| #1 \rangle}
\newcommand{\braket}[2]{\langle #1| #2\rangle}

\begin{document}

\title{Intrinsic CPT violation and decoherence \\ for entangled neutral mesons}
\date{\today}

\author{J. Bernab\'eu$^a$}
\author{N.E. Mavromatos$^{b}$}
\author{J. Papavassiliou$^a$}
\author{A. Waldron-Lauda$^b$}
\affiliation{$^a$Departamento de F\'\i sica Te\'orica and IFIC, Centro Mixto,
Universidad de Valencia-CSIC,
E-46100, Burjassot, Valencia, Spain. \\
$^b$King's College London, University of London, Department of Physics,
Strand WC2R 2LS, London, U.K.}

\begin{abstract}

We present a  combined treatment of quantum-gravity-induced decoherence
and intrinsic  CPT violation in entangled  neutral-Kaon states.
Our analysis takes  into consideration two  types of
effects: first, those  associated    with    the    loss    of
particle-antiparticle identity, as a  result of the ill-defined nature
of the CPT operator, and second, effects due to the non-unitary
evolution of the Kaons in  the space-time foam.  By studying a variety
of $\phi$-factory observables,  involving identical
as well as  general final states,
we derive  analytical expressions, to leading order  in the associated
CPT   violating   parameters,  for   double-decay   rates  and   their
time-integrated  counterparts.  Our  analysis shows  that  the various
types  of the  aforementioned effects  may be  disentangled  through
judicious combinations of appropriate observables in a $\phi$ factory.

\end{abstract}


\preprint{FTUV-05-0602}

\maketitle

\section{Introduction}

To date, the  Special Theory of Relativity, the  established theory of
flat  space-time  physics based  on  Lorentz  symmetry,  is very  well
tested. In  fact, during the current  year, it completes  a century of
enormous   success,   having  passed   very   stringent  and   diverse
experimental tests.  On  the other hand, a quantum  theory of Gravity,
that  is,  a  consistent   quantized  version  of  Einstein's  General
Relativity, still eludes us.  This  may be partially attributed to the
lack  of  any concrete  observational  evidence  on  the structure  of
space-time at  the characteristic scale  of quantum gravity  (QG), the
Planck mass-scale $M_P \sim 10^{19}$ GeV. One would  hope that, like any
other successful physical theory, a physically relevant quantum theory
of  gravity should lead  to experimental  predictions testable  in the
foreseeable future.  In the QG  case, however, such predictions may be
both numerically  suppressed and experimentally  difficult to isolate.
This is mainly  due to the extremely weak  nature of the gravitational
interaction as compared  with the rest of the  known forces in nature.
Specifically,  the dimension-full  coupling constant  of  gravity, the
Newton constant  $G_N=1/M_P^2$, appears  as a very  strong suppression
factor  of  any physical  observable  that  could  be associated  with
predictions   of  quantum   gravity.    If  the   feebleness  of   the
gravitational  interaction   were  to  be  combined   with  the  exact
conservation of all symmetries and properties in particle physics, one
would arrive at the inescapable conclusion that a true ``phenomenology
of QG'' should be considered wishful thinking.

In recent years, however, physicists  have ventured into the idea that
laws,   such  as   Lorentz   invariance  and   unitarity,  which   are
characteristic of  a flat  space-time quantum field  theory, \emph{may
not} be  valid in a  full \emph{quantum} theory of  curved space-time.
For  instance, the  invariance of  the laws  of Nature  under  the CPT
symmetry~\cite{cpt} (i.e.   the combined action  of charge conjugation
(C), parity-reflection (P), and time-reversal (T)) may not be exact in
the presence of QG effects.  Indeed, the possibility of a violation of
CPT invariance (CPTV) by QG has been raised in a number of theoretical
models,  that go  beyond  conventional local  quantum field  theoretic
treatments of gravity~\cite{hawking,ehns,lopez,peskin,kostel,bl,bmp}.

CPT violating  scenarios may be classified into  two major categories,
according to  the specific  way the symmetry  is violated.   The first
category   contains  models   in   which  CPTV   is  associated   with
(spontaneous)  breaking  of  (linear)  Lorentz  symmetry~\cite{kostel}
and/or  locality of  the  interactions~\cite{bl}. In  these cases  the
generator  of the CPT  symmetry is  a well-defined  quantum mechanical
operator, which  however \emph{does not commute}  with the (effective)
Hamiltonian  of the  system.  In  the  second major  category, the  CPT
operator, due to a variety of reasons, is \emph{ill-defined}. The most
notable example of  such a type of CPTV is due  to the entanglement of
the  matter theory  with an  ``environment'' of  quantum gravitational
degrees     of     freedom,      leading     to     decoherence     of
matter~\cite{ehns,lopez,peskin,bmp}.    This  decoherence,   in  turn,
implies,  by   means  of  a  powerful  mathematical   theorem  due  to
Wald~\cite{wald}, that the  quantum mechanical operator generating CPT
transformations cannot  be consistently defined.  Wald  refers to this
phenomenon as ``microscopic  time irreversibility'', which captures the
essence  of the  effect.  We  shall  refer to  such a  failure of  CPT
through   decoherence  as   ``intrinsic'',  in   contradistinction  to
``extrinsic''  violations,  related to  the  non-commutativity of  the
matter Hamiltonian with a well-defined CPT generator~\cite{kostel,bl}.

As  emphasized recently in  \cite{bmp}, this  intrinsic CPTV  leads to
modifications to  the concept  of an ``antiparticle''.   The resulting
loss  of particle-antiparticle  identity in  the  neutral-meson system
induces  a breaking of  the Einstein-Podolsky-Rosen  (EPR) correlation
imposed by Bose statistics.   Specifically, in the CPT invariant case,
the antiparticle of a given particle  is defined as the state of equal
mass (and life-time) and opposite  values for all charges.  If the CPT
operator is  well defined, such a  state is obtained by  the action of
this operator  on the corresponding particle state.   If, however, the
operator is  ill-defined, the particle and  antiparticle spaces should
be thought  of as  {\it independent} subspaces  of matter  states.  In
such  a case,  the usual  operating assumption  that  the electrically
neutral     meson    states    (Kaons     or    $B$     mesons)    are
\emph{indistinguishable}  from  (``identical''  to)  their  respective
antiparticles  (${\overline  K}^0$   or  ${\overline  B}^0$)  {\it  is
relaxed}. This, in turn, modifies the symmetry properties in the description
of (neutral) meson entangled states, and may bring about
deviations to their EPR correlations~\cite{bmp,bmp2}.
We emphasize, though, that the associated intrinsic
CPTV effects are treated as \emph{small perturbations},
being attributed to quantum-gravitational physics.
From now on, we will refer to this effect as the ``$\omega$ effect'',
due to the complex parameter $\omega$ used in its parametrization. 

It should be  stressed at this point that the  $\omega$ effect is {\it
distinct}, both  in its origin  as well as the  physical consequences,
from analogous effects generated by the evolution of the system in the
decoherence-inducing  medium of  QG, discussed  in  \cite{peskin}.  As
already mentioned  in \cite{bmp}, and  will be discussed in  detail in
the next  section, the $\omega$  effect appears immediately  after the
$\phi$-decay  (at time  $t=0$) due  to the  violation of  the symmetry
properties   of   the  entangled   Kaon   state  wave-function   under
permutations,  assuming conservation  of angular  momentum.   In other
words, the $\omega$ effect provides  an answer to the question: ``What
is the  (symmetry of the) initial  state of the  $K^0 {\overline K}^0$
system  in the presence  of a  QG-decoherence inducing  background ?''
This is  to be  contrasted to the  effects of  \cite{peskin}, stemming
from the  decoherent evolution of the  system in the  QG medium, which
have been interpreted as a violation of angular momentum conservation.
Our  working  hypothesis  is  that  the QG  medium  conserves  angular
momentum,  at  least as  far  as  the  $\omega$ effect  is  concerned,
extrapolating  from no-hair  theorems of  classical  macroscopic black
holes.

In \cite{bmp}  it was assumed  for simplicity that the  intrinsic CPTV
manifested itself  only at the  level of the  wave-function describing
the  entangled two-meson  state  immediately after  the  decay of  the
initial resonance, whereas the subsequent time evolution of the system
was taken to be due to an effective Hamiltonian. However, given that the origin
of this intrinsic CPTV is  usually attributed to the decoherent nature
of QG, in  the spirit discussed originally in  \cite{wald}, a complete
description  would  necessitate the  introduction  of such  ``medium''
effects in the time evolution as well.
To  accomplish  such  a  description,  at least  for  the  purpose  of
providing a phenomenological framework  for decoherent QG, no detailed
knowledge  of the  underlying  microscopic quantum  gravity theory  is
necessary. This  can 
be achieved  following the so-called  Lindblad or
mathematical  semi-groups   approach  to  decoherence~\cite{lindblad},
which  is a very  efficient way  of studying  open systems  in quantum
mechanics~\cite{footnote1}. The   
time  irreversibility  in  the   evolution  of  such
semigroups,  which  is  linked  to  decoherence, is  inherent  in  the
mathematical property of the lack of an inverse in the semigroup. This
approach  has   been  followed   for  the  study   of  quantum-gravity
decoherence in the case of neutral kaons~\cite{ehns,lopez}, as well as
other  probes, such  as  neutrinos~\cite{neutrinos,footnote2}.
The Lindblad approach to  decoherence does not
require any  detailed knowledge of the environment,  apart from energy
conservation,  entropy   increase  and  complete   positivity  of  the
(reduced)   density   matrix   $\rho(t)$   of  the   subsystem   under
consideration.  The basic evolution equation for the (reduced) density
matrix of  the subsystem in the  Lindblad approach is  {\it linear} in
$\rho(t)$.  We  notice that the Lindblad  part, expressing interaction
with  the ``environment''  cannot be  written  as a  commutator (of  a
Hamiltonian  function) with  $\rho$. Environmental  contributions that
can be cast in Hamiltonian evolution (commutator form) are absorbed in
an ``effective'' hamiltonian.

The purpose of this article is to
extend the analysis of \cite{bmp}, where the
intrinsic CPTV manifested itself only in the initial entangled meson state,
to include QG environmental entanglement, by resorting to the above-mentioned
Lindblad evolution formalism.
Specifically, we shall compute various observables relevant for precision
decoherence tests in a $\phi$-factory, placing emphasis on the possibility
of disentangling the decoherent evolution parameters $\alpha$, $\beta$, and  $\gamma$
of \cite{ehns} from the $\omega$ parameter of \cite{bmp}, expressing the
loss of particle-antiparticle identity mentioned above.
The structure of the article is as follows: in section 2 we present our
formalism, including notations and conventions, which agree with those used
in \cite{peskin}. In section 3 we describe the various observables
of a $\phi$ factory that constitute sensitive probes of our $\omega$ and
decoherence $\alpha,\beta,\gamma$ effects, and explain how the various
effects (or bounds thereof) can be disentangled experimentally.
In this section we restrict ourselves only to leading order
corrections in the CPTV and decoherence effects, and ignore $\epsilon '$
corrections. Such corrections do not affect the functional form of the CPTV and decoherent evolution terms of the various observables, and their inclusion is
discussed in an Appendix.
Conclusions and discussion
are presented in section 4. Particular emphasis in the discussion is
given in the contamination of the observable
giving traditionally (i.e. ignoring decoherent CPTV evolution)
the $\epsilon '$ effects,
by terms depending on the decoherence coefficients.
Finally, as already mentioned, in an Appendix we
give the complete formulae for the observables of section 3, including $\epsilon '$
corrections.

\section{Formalism}

In this section we shall follow the notation and conventions of \cite{dafne,peskin}.
The CP violating parameters  are defined as
 $\epsilon_S=\epsilon_M+\Delta$,
 $\epsilon_L=\epsilon_M-\Delta$,
 according to which the physical mass-eigenstates are expressed as
 \bea
|K_S\rangle &=&\frac{N_S}{\sqrt{2}}\left( (1 + \epsilon_S)|K_0\rangle +
(1 - \epsilon_S)|\overline{K_0}\rangle \right) \nonumber \\
|K_L\rangle &=& \frac{N_L}{\sqrt{2}}\left( (1 + \epsilon_L)|K_0\rangle -
(1 - \epsilon_L)|\overline{K_0}\rangle\right)~,
\eea
where $N_{S,L}$ are positive normalization constants,
$\epsilon_M $ is odd under CP, but even under CPT,
and $\Delta$ is odd under both CP and CPT.  Furthermore,
\bea
\epsilon_i^{\pm} &=&\epsilon_i \pm \frac{\beta}{d}
\nonumber \\
\epsilon_i^{\pm *} &=& \epsilon_i^* \pm \frac{\beta}{d^*},  \qquad i=L,S
\label{epsnqm}
\eea
with
\bea d &=& \Delta m +\frac{i}{2}\Delta \Gamma = |d|e^{i(\pi/2 -\phi_{SW})}
\nonumber \\
&=& \left((3.483 \pm 0.006)+ i (3.668 \pm 0.003)\right)\times 10^{-15}~{\rm GeV}
\eea
and the super-weak angle~\cite{Eidelman:2004wy} $\phi_{SW} = (43.5 \pm 0.7)^{\rm o}$.

To find the correct expressions for the
observables in the density matrix formalism of
\cite{ehns,lopez,peskin}, matching the standard phenomenology
in the CPT-conserving quantum mechanical case,
we use the following parametrization
for the amplitudes
of the decay
of $|K_1 \rangle,|K_2\rangle \to |\pi^+\pi^-\rangle$~\cite{peskin}:
\begin{eqnarray}
&& {\cal M}(K_1 \to \pi^+\pi^-) =\sqrt{2}A_0e^{i\delta_0} +
{\rm Re}A_2e^{i\delta_2} \nonumber \\
&& {\cal M}(K_2 \to \pi^+\pi^-) =i{\rm Im}A_2*e^{i\delta_2}
\label{amplitudes}
\end{eqnarray}
where $A_0$ is real, and
$\epsilon ' = \frac{i}{\sqrt{2}}\frac{{\rm Im}A_2}{A_0}e^{i\delta}$,
$\delta \equiv \delta_2 - \delta_0$, and
$|\eta_{+-}|^2 \equiv |\frac{{\cal M}(K_L \to \pi^+\pi^-)}{{\cal M}(K_S \to \pi^+\pi^-)}|^2 = |\epsilon_L
+\frac{1}{\sqrt{2}}\frac{i{\rm Im}A_2}{A_0}e^{i\delta}|^2$,
$|\eta_{00}|^2 \equiv |\frac{{\cal M}(K_L \to \pi^0\pi^0)}{{\cal M}(K_S \to \pi^0\pi^0)}|^2 = |\epsilon_L
-\sqrt{2}\frac{i{\rm Im}A_2}{A_0}e^{i\delta}|^2$.
We also introduce the standard parametrization
$\eta_{+-} =  \langle \pi^+\pi^-|K_L\rangle/\langle \pi^+\pi^-|K_S\rangle =
|\eta_{+-}| e^{i\phi_{+-}}$, where~\cite{Eidelman:2004wy}
$|\eta_{+-}|= (2.288 \pm 0.014)\times 10^{-3}$,\,
$\phi_{+-} = (43.4 \pm 0.7)^{\rm o}$. Similar parametrizations can be used
for the corresponding rations representing Kaon decays to two neutral
as well as three pions. The corresponding phases are denoted $\phi_{00}$ and
$\phi_{3\pi}$.
We remind the reader that within quantum
mechanics $\eta_{+-} = \epsilon_L + \epsilon'$, $\eta_{00}=
\epsilon_L - 2\epsilon'$.

All the above numbers are extracted without assuming
CPT invariance.
In the presence of CPTV
decoherence parameters ($\beta$) this relation is modified,
given that now $\epsilon_{L,S} $ are replaced by $\epsilon_{L,S}^{\pm}$ (\ref{epsnqm}). In that case~\cite{peskin}, $\epsilon_L^{-}$
is related to $\overline \eta_{+-}$, which replaces the quantum-mechanical
$\eta_{+-}$ defined above, through
$|\overline \eta_{+-}|e^{i\phi_{+-}} = \epsilon_L^{-} + Y_{+-}$, where
$Y_{+-}$ includes the $\epsilon '$ effects,
$Y_{+-} = \langle \pi^+\pi^-|K_2\rangle/\langle \pi^+\pi^-|K_1\rangle$, with
 $K_{1} = \frac{1}{\sqrt{2}}\left( |K^0\rangle + |{\overline K}^0\rangle \right)$ and
$K_{2} = \frac{1}{\sqrt{2}}\left( |K^0\rangle - |{\overline K}^0\rangle\right)$
the standard CP eigenstates.
We will also use the abbreviations
$\Delta \epsilon_{SL}= \epsilon_S -\epsilon_{L}$ and
 $\sum \epsilon_{SL}= \epsilon_S +\epsilon_{L}$,
and $\overline{\Gamma} = \frac{1}{2}(\Gamma_L+\Gamma_S)$, as well as the definition~\cite{peskin}
\bea
R_L &=&  |\epsilon_L^-|^2 + \frac{\gamma}{\Delta \Gamma} + 4\frac{\beta}{\Delta \Gamma}{\rm Im}
\left(\frac{\epsilon_L^-d}{d^*}\right),
\eea
which we shall make use of in the present article. We also note the current experimental
values $\sqrt{R_L} = (2.30 \pm 0.035)\times 10^{-3}$, $2 {\rm Re}\epsilon_L^+ =
(3.27 \pm 0.12)\times 10^{-3}$.
Finally, the relevant density matrices read
$\rho_L = \ket{K_L}\bra{K_L}$, $\rho_S = \ket{K_S}\bra{K_S}$, $\rho_I = \ket{K_S}\bra{K_L}$,
$\rho_{\bar{I}} = \ket{K_L}\bra{K_S}$.
In the present article we shall use values for the
decoherence parameters
$\alpha$, $\beta$, and $\gamma$, which saturate
the experimental bounds
obtained by the CPLEAR experiment~\cite{cplear}:
\be
\alpha < 4.0 \times 10^{-17}~{\rm GeV} ,\qquad |\beta| < 2.3 \times 10^{-19}~{\rm GeV} ,\qquad
\gamma < 3.7 \times 10^{-21}~{\rm GeV}
\label{cplearbound}
\ee
We note at this stage that in certain models 
\cite{benatti}, where complete positivity 
of the density matrix of the entangled kaon state
has been invoked, there is only one non-vanishing decoherence 
parameter ($\alpha = \gamma$) 
in the formalism of \cite{ehns}, which we adopt here.
However, as the analysis of \cite{benatti} has demonstrated, 
in such completely positive entangled 
models
there are other parametrizations, with more decoherence parameters.
Our intrinsic CPT-violation 
$\omega$-effects~\cite{bmp} can be easily incorporated
in those formalisms, which however we shall not follow here.
For our purposes, complete positivity is 
achieved as the special case of only one decoherence parameter 
($\gamma = \alpha $) in the formulae that follow. We emphasize, however, that
the issue of complete positivity is not entirely clear
in a quantum-gravity context, where the relevant master equation
may even be non-linear~\cite{emnnl}.    

In conventional
formulations of {\it entangled} meson
states~\cite{dunietz,botella,bernabeu}
one imposes the requirement of {\it Bose statistics}
for the state $K^0 {\overline K}^0$ (or $B^0 {\overline B}^0$),
which implies that the physical neutral meson-antimeson state
must be {\it symmetric} under the combined operation $C{\cal P}$,
with $C$ the charge conjugation and
${\cal P}$ the operator that permutes the spatial coordinates.
Specifically, assuming
{\it conservation} of angular momentum, and
a proper existence of the {\it antiparticle state} (denoted by a bar),
one observes that,
for $K^0 {\overline K}^0$ states which are $C$-conjugates with
$C=(-1)^\ell$ (with $\ell$ the angular momentum quantum number),
the system has to be an eigenstate of
${\cal P}$ with eigenvalue $(-1)^\ell$.
Hence, for $\ell =1$, we have that $C=-$, implying ${\cal P}=-$.
As a consequence
of Bose statistics this ensures that for $\ell = 1$
the state of two identical bosons is forbidden~\cite{dunietz}.
However,  these
assumptions may not be valid if CPT symmetry is intrinsically violated,
in the sense of the loss of particle-antiparticle identity.
In such a case
${\overline K}^0$ cannot be considered
as identical to ${K}^0$, and thus the requirement of $C {\cal P} = +$, imposed
by Bose-statistics, is relaxed.
Therefore, the initial state
after the $\phi$ decay, when expressed in terms of mass-eigenstates
contains,
in addition to the standard   $K_L K_S$ terms,
otherwise forbidden terms of the type $K_L K_L$, $K_S K_S$:
\begin{eqnarray}
|i\rangle &=&
C \bigg[ \left(|K_S({\vec k}),K_L(-{\vec k})\rangle
- |K_L({\vec k}),K_S(-{\vec k})\rangle \right)\nonumber \\
&+& \omega \left(|K_S({\vec k}), K_S(-{\vec k})\rangle
- |K_L({\vec k}),K_L(-{\vec k})\rangle \right)\bigg]
\label{bph}
\end{eqnarray}
where C is to be computed, and $\omega$ is a complex parameter.
We hasten to emphasize at this stage that the genuine, quantum
gravity induced $\omega$-effect, 
due to the loss of particle-antiparticle identity,  
{\it should not} be confused with the 
ordinary C-even-background effects that 
have been studied in \cite{benatti}. In fact the $\omega$-effect 
can be easily disentangled from background effects, as 
has been discussed in \cite{bmp}.

Under the non-unitary decoherent
Lindblad evolution~\cite{lindblad}, appropriately tailor to
the present problem~\cite{lopez,peskin}
\be
\partial_t \rho (t) = i\rho(t)H - iH^\dagger\rho(t) + \widehat{\delta H}\rho(t)
\label{lindblad}
\ee
where $H$ denotes the (non-hermitian) Hamiltonian of the system
(taking decay into account) and $\widehat{\delta H}$
contains the decoherent effects,
the initial state evolves to a mixed state, which
assumes the general structure \cite{peskin}
\be
\rho =  \sum_{i,j} A_{ij} \rho_i\otimes \rho_j~,\qquad i,j=S,L~,
\ee
where the constants  $A_{ij}$ depend on the decoherence parameters.

The relevant observables are computed by means of double-decay rates with the following
generic structure
 \bea
    \mathcal{P}(f_1,\tau_1;f_2, \tau_2)
    =\sum_{ij}A_{ij}tr[\rho_i \mathcal{O}_{f_1}]
    tr[\rho_j \mathcal{O}_{f_2}]e^{-\lambda_i\tau_1-\lambda_j
    \tau_2},
\label{ddr}
\eea
where the constants  $\lambda_i$, $\lambda_j$ also depend on the decoherence parameters.

The double decay rate $\mathcal{P}(f_1,\tau ;f_2, \tau )$ interpolated
at  \emph{equal times} $\tau_1  = \tau_2  = \tau$  is a  quantity that
pronounces the  unusual time dependences of  the decoherent evolution,
which will also be calculated in this work.

Finally,
of particular experimental interest are also integrated distributions
at fixed time intervals $\Delta \tau = \tau_2 - \tau_1 > 0$ (which we can
assume for our purposes here):
 \bea
    \mathcal{{\overline P}}(f_1;f_2; \Delta\tau > 0 )
    = \frac{1}{2}\int_{\Delta \tau }^\infty d(\tau_1 + \tau_2)
\mathcal{P}(f_1,\tau_1;f_2,\tau_2)~.
\label{iddr}
\eea
where the factor $\frac{1}{2}$ originates from the Jacobian when changing variables
$(\tau_1,\tau_2 )\to (\tau_1 +\tau_2, \Delta \tau)$.

In the particular case of a  $\phi$ factory that we analyze here,
the pertinent density matrix describing the decay of one kaon of momentum ${\vec k}$ at time $\tau_1$
to a final state $f_1$ and the other of momentum ${-\vec k}$ at time $\tau_2$
to a final state $f_2$ is given  by
 \bea
    \rho &=& \rho_S\otimes \rho_L\left(1+\omega\Delta\epsilon_{SL}
    +\omega^*\Delta\epsilon^*_{SL}\right) e^{-\Gamma_L \tau_2
    -\Gamma_S \tau_1}+ \rho_L\otimes \rho_S \left(1-\omega\Delta\epsilon_{SL}
    -\omega^*\Delta\epsilon^*_{SL}\right)
    e^{-\Gamma_L \tau_1-\Gamma_S \tau_2}
    \nonumber\\&&-\left(\rho_I\otimes
    \rho_{\bar{I}}(1+\omega\Delta\epsilon_{SL}-\omega^*\sum\epsilon^*_{SL})
    e^{-i\Delta m (\tau_1-\tau_2)}\right.
    \nonumber\\&& \left.+ \rho_{\bar{I}}\otimes \rho_I
    (1-\omega\Delta\epsilon_{SL}+\omega^*\sum\epsilon^*_{SL})e^{+i\Delta m
    (\tau_1-\tau_2)}\right)e^{-(\overline{\Gamma}+\alpha-\gamma)(\tau_1+\tau_2)}
    \nonumber\\&& -\left(\left(|\omega|^2 - \frac{i\alpha}{\Delta m}\right)
    \rho_I \otimes \rho_I e^{-i\Delta m
    (\tau_1+\tau_2)}  + \left(|\omega|^2 + \frac{i\alpha}{\Delta m}\right)
\rho_{\bar{I}} \otimes
    \rho_{\bar{I}}e^{+i\Delta m (\tau_1+\tau_2)}\right)e^{-(\overline{\Gamma}+\alpha-\gamma)(\tau_1+\tau_2) }
    \nonumber\\&& +\left(|\omega|^2- \frac{2\gamma}{\Delta \Gamma}\right)
    \rho_S \otimes \rho_S e^{-\Gamma_S (\tau_1+\tau_2)} +
     \left(|\omega|^2 + \frac{2\gamma}{\Delta \Gamma}\right)\rho_L \otimes \rho_L
    e^{-\Gamma_L (\tau_1+\tau_2)}
    \nonumber\\&& +\rho_I\otimes \rho_S e^{- i\Delta m \tau_1 -\Gamma_S
    \tau_2 -(\overline{\Gamma}+\alpha-\gamma) \tau_1}\left(\omega
     -\frac{2\beta}{d}\right)
      -\rho_S\otimes \rho_I e^{- i\Delta m \tau_2 -\Gamma_S
    \tau_1 -(\overline{\Gamma}+\alpha-\gamma)\tau_2} \left(\omega
     +\frac{2\beta}{d}\right)
    \nonumber\\&& +\rho_{\bar{I}}\otimes \rho_S e^{+i\Delta m
    \tau_1 -\Gamma_S \tau_2 -(\overline{\Gamma}+\alpha-\gamma)\tau_1}\left(\omega^*
     -\frac{2\beta}{d^*}\right)
      -\rho_S\otimes \rho_{\bar{I}} e^{+ i\Delta m \tau_2 -\Gamma_S
    \tau_1 -(\overline{\Gamma}+\alpha-\gamma)\tau_2} \left(\omega^*
    +\frac{2\beta}{d^*}\right)
    \nonumber\\&& +\rho_I\otimes \rho_L e^{-i\Delta m \tau_1 -\Gamma_L \tau_2
    -(\overline{\Gamma}+\alpha-\gamma)\tau_1}\left(\omega^*
     +\frac{2\beta}{d^*}\right)
     -\rho_L\otimes
    \rho_I e^{-i\Delta m \tau_2 -\Gamma_L \tau_1
    -(\overline{\Gamma}+\alpha-\gamma)\tau_2}\left(\omega^*
    -\frac{2\beta}{d^*}\right)
    \nonumber\\&& +\rho_{\bar{I}}\otimes \rho_L e^{ +i\Delta m \tau_1 -\Gamma_L \tau_2
    -(\overline{\Gamma}+\alpha-\gamma)\tau_1}\left(\omega
     +\frac{2\beta}{d}\right)
     -\rho_L\otimes
    \rho_{\bar{I}}e^{+i\Delta m \tau_2 -\Gamma_L \tau_1
    -(\overline{\Gamma}+\alpha-\gamma)\tau_2}\left(\omega
    -\frac{2\beta}{d}\right)\nonumber\\&&
      {}
\label{initden}
\eea
from the  which  the  expression for  the  various constants  in
Eq.(\ref{ddr}) may be  gleaned.  An important remark is  now in order.
Notice that  above we have treated  $|\omega |^2$ effects  as being of
comparable order  to decoherence $\alpha, \gamma$, and  have kept only
terms at  most linear  in the decoherence  parameters. This is  a very
simplifying assumption, which,  due to a lack of  a microscopic theory
of QG, cannot be made rigorous.  However, as we shall discuss below, a
reasonable \emph{a posteriori} justification of this approximation may
come  from the  fact that  such terms  appear on  an equal  footing as
medium-generated corrections to  certain physically important terms of
(\ref{initden}) to be analyzed below.

In Eq.(\ref{initden})  we notice that,  starting from the  third line,
terms of entirely novel type make their appearance. Such terms are due
to  both the  intrinsic  CPTV loss  of particle-antiparticle  identity
($\omega$-related)          and          decoherent          evolution
($\alpha$,$\beta$,$\gamma$,-related).  Such terms persist in the limit
$\tau_1 = \tau_2 = 0$, due to a specific choice of boundary conditions
in the  solution of the appropriate  density-matrix evolution equation
(\ref{lindblad}),  expressing the  omnipresence of  the  QG space-time
foam  effects. In  particular,  terms linear  in  $\omega$ are  always
accompanied  by  the parameter  $\beta$,  whereas  terms quadratic  in
$\omega$ always combine with  the $\alpha$ or $\gamma$ parameters.  Of
particular  physical  importance  are  the otherwise  forbidden  terms
$\rho_L \otimes \rho_L$ and  $\rho_S \otimes \rho_S$.  At first sight,
it is  tempting to interpret~\cite{peskin}  such terms as  signaling a
subtle breakdown of angular  momentum conservation, in addition to the
CPTV introduced  by the ``medium''.   That seems to be  an inescapable
conclusion,  if one were  to begin  the time  evolution from  a purely
antisymmetric wave function (i.e.  with $\omega=0$ in (\ref{bph})), as
assumed  in~\cite{peskin},  which  is  an eigenstate  of  the  orbital
angular momentum  $L$ with  eigenvalue $\ell =  1$.  Since,  then, the
medium  generates in the  final state  forbidden terms  violating this
last   property,   one   interprets   this  as   non-conservation   of
$L$.  However,  if  one  accepts~\cite{bmp} that  the  intrinsic  CPTV
affects  not  only   the  evolution  but  also  the   concept  of  the
\emph{antiparticle},  such terms  are present  already in  the initial
state.   In such  a case,  there is  no issue  of  non-conservation of
angular momentum; the  only effect of the time  evolution is to simply
modify  the  relative  weights  between $\rho_L  \otimes  \rho_L$  and
$\rho_S \otimes\rho_S$ terms.

One  can  therefore   say  that  QG  may  behave   as  a  CPTV  medium
differentiating between particles  and antiparticles.  From a physical
point of view  one may even draw a vague analogy  between QG media and
``regenerators'', in  the following sense: exactly  as the regenerator
differentiates  between  $K^0$   and  $\overline{K}^0$,  by  means  of
different relative interactions, in  a similar way QG acts differently
on particles and antiparticles;  however, unlike the regenerator case,
where  the experimentalist  can  control its  position,  due to  their
universal nature,  QG effects  are present even  in the  initial state
after the $\phi$ decay.

\section{Observables at  $\phi$ factories}

In this section we present a detailed study of a variety of observables
measurable at $\phi$ factories, and determine their dependence on
both intrinsic CPTV ($\omega$) and decoherence ($\alpha$, $\beta$, $\gamma$)
parameters, in an attempt to disentangle and separately constrain the
two possible effects. Specifically, we will derive expressions for
the double-decay rates of Eq.(\ref{ddr})
and their integrated counter-parts of  Eq.(\ref{iddr}), for the cases of
identical as well as general (non-identical) final states.
We will restrict ourselves to linear effects in $\alpha$, $\beta$, $\gamma$,
while keeping terms linear and quadratic in $\omega$, for the reasons explained
in the previous section.
For our purposes here we shall be interested in the decays
of $K_{L,S}$ to final states consisting of two pions, $\pi^+\pi^-$ or
$\pi^0$, three pions, as well as semileptonic decays to $\ell^+ \pi^- \nu $
or $\ell^- \pi^+ {\overline \nu}$. In this section we shall
also ignore explicit $\epsilon '$ effects. Such effects appear in
the branching ratios $\eta$ of various pion channels, and
their inclusion affects the form of the associated observables.
We present complete formulae for the relevant double-decay time distributions,
including such effects,  in an Appendix. As we note there,
the inclusion of such effects does not affect significantly
the functional form of the decoherent/CPTV evolution.

In the density-matrix formalism of \cite{ehns,peskin}
the relevant observables are given by
\begin{eqnarray}
&& {\cal O}_{+-} = \left(\begin{array}{cc} {\cal X}_{+-}
\qquad \quad \quad {\cal Y}_{+-}
\\ {\cal Y}_{+-}^*
\quad |{\rm Im}A_2e^{i\delta}|^2
\end{array}\right) \nonumber \\
&&{\cal O}_{00} = \left(\begin{array}{cc} {\cal X}_{00}
\qquad \quad \quad \qquad \quad \quad {\cal Y}_{00}\\
{\cal Y}_{00}^* \quad \quad \quad |-2i{\rm Im}A_2e^{i\delta}|^2
\end{array}\right) \nonumber \\
&& {\cal O}_{\ell^+} =\frac{|a|^2}{2}
\left(\begin{array}{cc} 1 \quad 1 \\1 \quad 1 \end{array}\right)~,
\nonumber \\
&& {\cal O}_{\ell^-} =\frac{|a|^2}{2}
\left(\begin{array}{cc} 1 \quad -1 \\-1 \quad 1 \end{array}\right)~,
\nonumber \\
&& \mathcal{O}_{3\pi}=|X_{3\pi}|^2\left(%
\begin{array}{cc}
  0 & 0 \\
  0 & 1 \\
\end{array}%
\right)~.
\label{obsfnal}
\end{eqnarray}
where ${\cal X}_{+-} = |\sqrt{2}A_0 +
{\rm Re}A_2e^{i\delta}|^2 $,
${\cal Y}_{+-} = (\sqrt{2}A_0 + {\rm Re}A_2e^{-i\delta})
(i{\rm Im}A_2e^{i\delta})$,
${\cal X}_{00} = |\sqrt{2}A_0 -
2{\rm Re}A_2e^{i\delta}|^2 $ and
${\cal Y}_{00} = (\sqrt{2}A_0 - 2{\rm Re}A_2e^{-i\delta})
(-2i{\rm Im}A_2e^{i\delta})$, and
$X_{3\pi}=\braket{3\pi}{K_2}$,
in an obvious shorthand notation $O_f$,
where $f$ denotes the observable associated
with a Kaon decay leading to $f$ final state,
for instance $f=+-$ denotes decay to $\pi^+\pi^-$,
$f=00$ denotes decay to two $\pi^0$, $f=\ell^+$ denotes
semileptonic decay $\ell^+ \pi^- \nu $ {\it etc.}
Notice that if $\epsilon '$ effects
are ignored,
which we shall do here for brevity,
then we do not distinguish
between the observables for the two cases of the
two-pion final states.

Note that above we ignored CPT violating effects
in the decay amplitudes to the final states. If such effects are included
then the observables acquire the form~\cite{peskin}:
\begin{eqnarray}
&& {\cal O}_{+-} = |X_{+-}|^2
\left(\begin{array}{cc} 1
\qquad Y_{+-}\\
Y_{+-}^*
\quad
|Y_{+-}|^2 \end{array}\right)~ \nonumber \\
&&{\cal O}_{00} = |X_{00}|^2
\left(\begin{array}{cc} 1
\qquad Y_{00}\\
Y_{00}^*
\quad
|Y_{00}|^2 \end{array}\right)~ \nonumber \\
&& {\cal O}_{\ell^+} =\frac{|a + b|^2}{2}
\left(\begin{array}{cc} 1 \quad 1 \\1 \quad 1 \end{array}\right)~,
\nonumber \\
&&{\cal O}_{\ell^-} =\frac{|a - b|^2}{2}
\left(\begin{array}{cc} 1 \quad -1 \\-1 \quad 1 \end{array}\right)~,
\nonumber \\
&& \mathcal{O}_{3\pi}=|X_{3\pi}|^2\left(%
\begin{array}{cc}
  |Y_{3\pi}|^2 & Y_{3\pi}^* \\
  Y_{3\pi} & 1 \\
\end{array}%
\right)~.
\label{obsfnalcptv}
\end{eqnarray}
where $X = <\pi\pi|K_+>$, $Y = \frac{<\pi\pi|K_->}{<\pi\pi|K_+>}$,
and more specifically $Y_{+-} = (\frac{{\rm Re}(B_0)}{A_0}) + \epsilon '$,
$Y_{+-} = (\frac{{\rm Re}B_0}{A_0}) -2\epsilon '$,
$Y_{3\pi}=\frac{\braket{3\pi}{K_1}}{\braket{3\pi}{K_2}}$,
and ${\rm Re}(a/b)$ and ${\rm Re}(B_0)/A_0$
parametrize CPT violation in  the decay
amplitudes to the appropriate final states.

\subsection{Identical final states}

Consider first the case where the kaons decay to $\pi^+\pi^-$.
Ignoring CPTV in the decays and $\epsilon '$ effects,
the relevant observable, $\mathcal{O}_{+-}$, is
obtained from the first of
Eq.(\ref{obsfnal}) by setting $A_2 =0$, namely
 \ban
    \mathcal{O}_{+-}=\left(%
    \begin{array}{cc}
  2A_0^2 & 0 \\
  0 & 0 \\
\end{array}%
\right)
 \ean
The corresponding double-decay rate is then
 \bea
    &\mathcal{P}&(\pi^+ \pi^-,\tau_1;\pi^+ \pi^-,\tau_2)=
    2A_0^4\left[ R_L(
    e^{-\Gamma_L \tau_2 -\Gamma_S \tau_1} +
    e^{-\Gamma_L \tau_1-\Gamma_S \tau_2})\right.
    \nonumber \\ && -2|\overline{\eta}_{+-}|^2\cos(\Delta m
    (\tau_1-\tau_2)) e^{-(\overline{\Gamma}+\alpha -\gamma)(\tau_1+\tau_2)}
   +\left(|\omega|^2-\frac{2\gamma}{\Delta \Gamma} -\frac{4\beta}{|d|}
   |\overline{\eta}_{+-}|\frac{\sin\phi_{+-}}{\cos\phi_{SW}}\right)
    e^{-\Gamma_S (\tau_1+\tau_2)}
    \nonumber\\ && +\frac{4\beta |\overline{\eta}_{+-}|}{|d|}
    \sin(\Delta m \tau_1 +\phi_{+-}-\phi_{SW})e^{-(\overline{\Gamma}+\alpha
    -\gamma)\tau_1-\Gamma_S\tau_2}
    \nonumber\\ && +\frac{4\beta |\overline{\eta}_{+-}|}{|d|}
    \sin(\Delta m \tau_2 +\phi_{+-}-\phi_{SW})e^{-(\overline{\Gamma}+\alpha -\gamma)
    \tau_2-\Gamma_S\tau_1}
    \nonumber\\ && +2|\omega||\overline{\eta}_{+-}|(\cos(\tau_1+\phi_{+-}-2\Omega)
    e^{-\Gamma_S(\tau_2)-(\overline{\Gamma}+\alpha-\gamma)\tau_1}
   \left. -\cos(\tau_2+\phi_{+-}-\Omega)
    e^{-\Gamma_S\tau_1-(\overline{\Gamma}+\alpha-\gamma)\tau_2})
    \right]\nonumber \\ &&
    \label{twopion}
 \eea

\begin{figure}[tb]
\includegraphics[width=8cm]{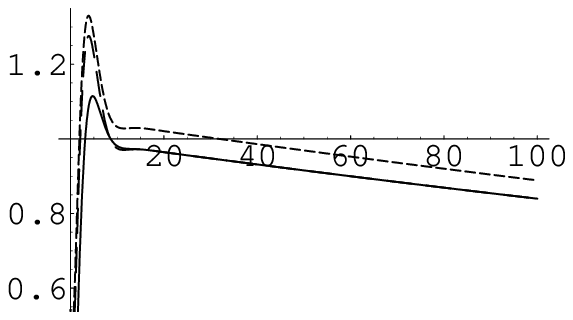}
\includegraphics[width=8cm]{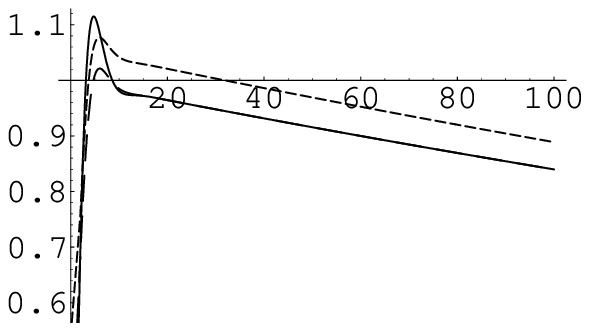}
\includegraphics[width=8cm]{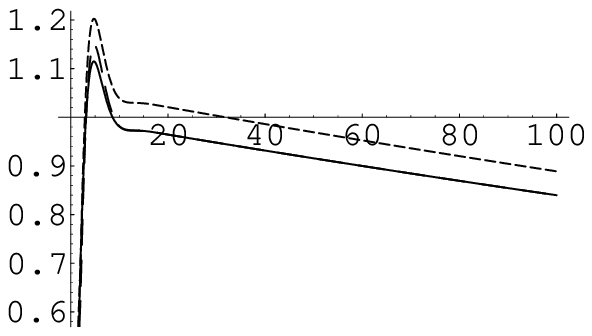}
\includegraphics[width=8cm]{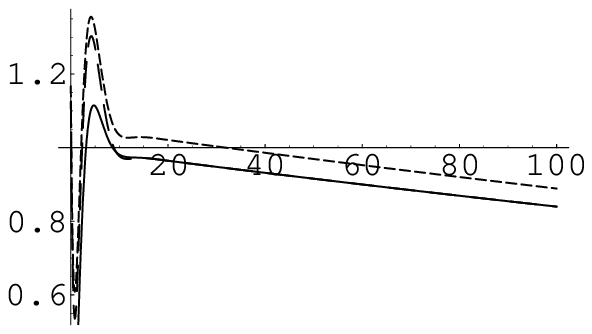}
\caption{{\it Plots of
$\overline{\mathcal{P}}(\pi^{\pm},\pi^{\pm},\Delta\tau)$ vs.
$\Delta \tau$ extended to a large range of $\Delta \tau$. The
dashed curves are with $\alpha,\beta,\gamma,\omega$ all non-zero,
the long-dashed curve has only $\omega \ne 0$, whilst the solid
curve represents $\alpha,\beta,\gamma,\omega=0$. Here and in all
subsequent figures the values for $\alpha$,$\beta$, and $\gamma$
saturate the CPLEAR bound of Eq.(\ref{cplearbound}). The values
for $\omega$ and $\Omega$ are (from left to right, and top to
bottom) i) $|\omega|=|\overline{\eta}_{+-}|$,
$\Omega=\phi_{+-}-0.16\pi$, ii) $|\omega|=|\overline{\eta}_{+-}|$,
$\Omega=\phi_{+-}+0.95\pi$, iii)
$|\omega|=0.5|\overline{\eta}_{+-}|$, $\Omega=\phi_{+-}+0.16\pi$,
iv) $|\omega|=1.5|\overline{\eta}_{+-}|$, $\Omega=\phi_{+-}$.
$\bar{\mathcal{P}}$ is in units of $|\overline{\eta}_{+-}|^2
\tau_s 4 A_0^{4}$ and $\Delta\tau$ is in units of $\tau_S$.}}
\label{pipmextended}
\end{figure}

\begin{figure}[tb]
\includegraphics[width=8cm]{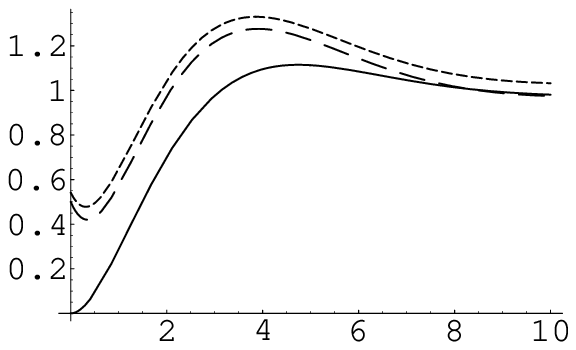}
\includegraphics[width=8cm]{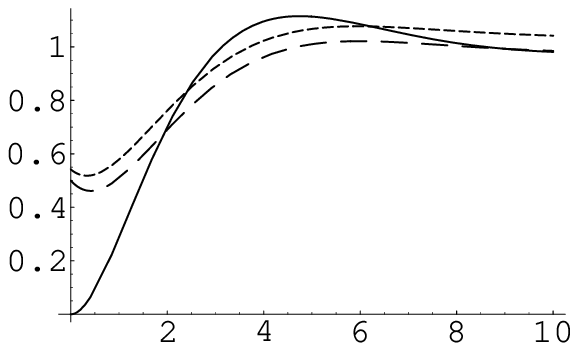}
\includegraphics[width=8cm]{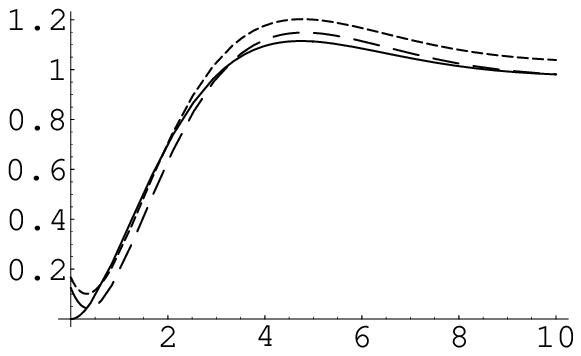}
\includegraphics[width=8cm]{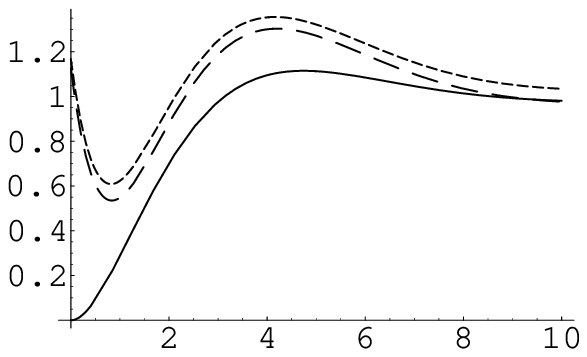}
\caption{{\it As in Fig. \ref{pipmextended}, but shown in detail
for small values of $\Delta \tau $ near the origin, to illustrate
the difference between the various cases. Again, the dashed curves
are with $\alpha,\beta,\gamma,\omega$ all non-zero, the
long-dashed curve has only $\omega \ne 0$, whilst the solid curve
represents $\alpha,\beta,\gamma,\omega=0$.}} \label{pipm}
\end{figure}

The integrated observable $\overline{\mathcal{P}}(f_1,f_2,\Delta \tau)$  is obtained
by integrating over $(\tau_1+\tau_2)$ keeping $\Delta \tau$ fixed. The result is:
 \bea
    &\overline{\mathcal{P}}&(\pi^+ \pi^-;\pi^+ \pi^-;\Delta\tau)=
    A_0^4\left. \bigg[ R_L \frac{e^{-\Delta \tau \Gamma_L}+e^{-\Delta \tau \Gamma_S}}
    {\Gamma_L+\Gamma_S}\right.
   \nonumber \\ &&\left. -|\overline{\eta}_{+-}|^2 \cos(\Delta m\Delta
    \tau)\frac{e^{-(\overline{\Gamma}+\alpha -\gamma)\Delta\tau}}
    {(\overline{\Gamma}+\alpha -\gamma)}
    +\left(|\omega|^2-\frac{2\gamma}{\Delta \Gamma}
    -\frac{4\beta}{|d|}
   |\overline{\eta}_{+-}|\frac{\sin\phi_{+-}}{\cos\phi_{SW}} \right)
    \frac{ e^{-\Gamma_S \Delta\tau}}{2\Gamma_S}\right.
    \nonumber \\ &&\left. + \frac{1}{\Delta m^2+(\overline{\Gamma}+\alpha-\gamma)^2
    +2(\overline{\Gamma}+\alpha-\gamma)\Gamma_S+\Gamma_S^2}\right.
   \nonumber\\ &&\left. \times \left. \bigg[ \frac{4\beta
    |\overline{\eta}_{+-}|}{|d|}\left((\Delta m\cos(\phi_{+-}-\phi_{SW})
    +(\overline{\Gamma}+\alpha-\gamma+\Gamma_S)\sin(\phi_{+-}-\phi_{SW}))
    e^{-\Delta\tau \Gamma_S}
    \right.\right.\right.
   \nonumber \\ && \left. +e^{-\Delta\tau(\overline{\Gamma}+\alpha-\gamma)}
    (\Delta m\cos(\Delta m\Delta\tau +\phi_{+-}-\phi_{SW})\right.
   \nonumber \\ && \left.\left.+(\overline{\Gamma}+\alpha-\gamma+\Gamma_S)
    \sin(\Delta m\Delta\tau +\phi_{+-}-\phi_{SW}))
    \right)\right.
    \nonumber\\ && \left. +2|\omega||\overline{\eta}_{+-}|e^{-\Delta\tau \Gamma_S}
    ((\overline{\Gamma}+\alpha-\gamma+\Gamma_S)\cos(\phi_{+-}-\Omega)
    -\Delta m \sin(\phi_{+-}-\Omega))\right.
    \nonumber \\ && \left. -2 |\omega||\overline{\eta}_{+-}| e^{(\overline{\Gamma}+\alpha-\gamma)\Delta\tau}
    ((\overline{\Gamma}+\alpha-\gamma+\Gamma_S)\cos(\Delta m\Delta\tau+\phi_{+-}-\Omega )\right.
   \nonumber \\
&& \left. \left.-\Delta m\sin(\Delta m\Delta\tau+\phi_{+-}-\Omega))\right. \bigg]\right. \bigg]
\nonumber \\ &&
\label{twopionint}
\eea
which we plot in Figs. \ref{pipmextended}, \ref{pipm}.
In the figures we assume rather large vales of $|\omega |$,
of  order $|\overline{\eta}_{+-}|$.
As evidenced by looking at the curves for relatively
large values of $\Delta \tau$, this assumption, together with the
assumed values of the decoherence parameters (\ref{cplearbound}),
allows for a rather clear disentanglement of the decoherence plus
$\omega$ situation from that with only $\omega$ effects present (i.e.
unitary evolution). Indeed, in the latter case,
the result for the pertinent
time-integrated double-decay rate
quickly converges to the quantum-mechanical situation, in contrast to the
decoherent case. Moreover, the non-zero value of the time-integrated
asymmetries at $\Delta \tau = 0$, is another clear deviation from the
quantum mechanical case, which in the (unitary evolution)
case with only $\omega \ne 0$ is pronounced, due to the presence of
$|\overline{\eta}_{+-} | $ factors~\cite{bmp}.

We notice at this stage, though, that this may not be the case
in an actual quantum gravity foam situation.
For instance, it has been argued~\cite{benatti}
that if \emph{complete positivity}
is imposed for the density matrix of entangled states, then
the only non-trivial decoherence parameter would be $\alpha = \gamma > 0$,
whilst $\beta =0$.
It is not inconceivable
that $|\omega |^2$
are of the same order as $2\gamma/\Delta \Gamma$,
in which case,
for all practical purposes the integrated curve
for the two-pion double decay rate (\ref{twopionint})
would pass through zero for $\Delta \tau = 0$.
In such a situation, one should \emph{also} look
at the linear in
$\omega$ interference terms for $\Delta \tau \ne 0$, in order to
disentangle the $\omega$ from $\gamma$ effects.
Lacking a fundamental microscopic theory underlying
these phenomenological considerations, however, which would
make definite predictions on the relative order of the various
CPTV effects, we are unable to make more definite statements at present.

\begin{figure}[tb]
\includegraphics[width=8cm]{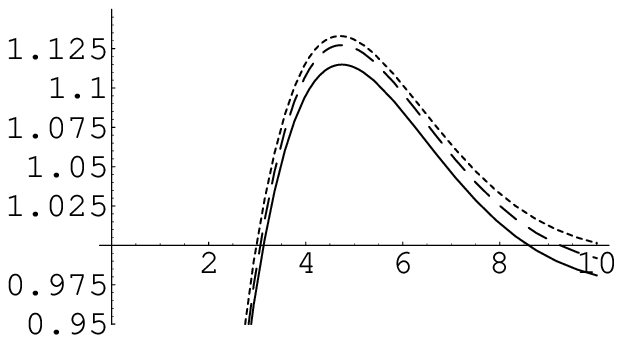}
\includegraphics[width=8cm]{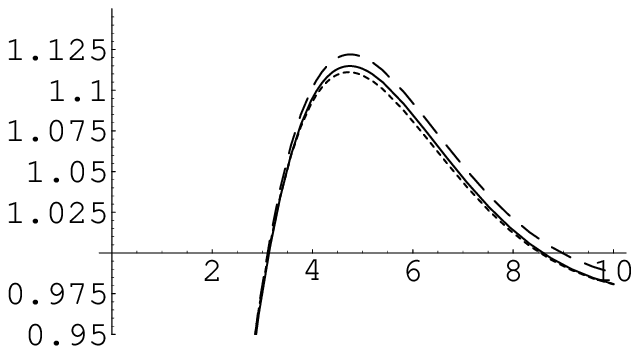}
\includegraphics[width=8cm]{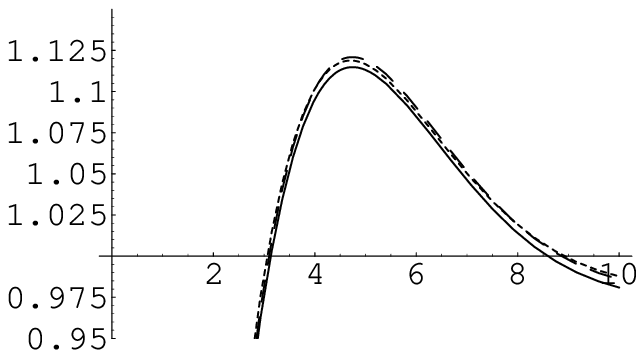}
\includegraphics[width=8cm]{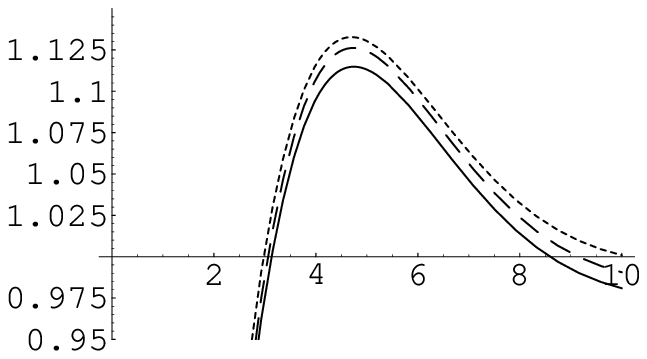}
\caption{{\it Plots of
$\overline{\mathcal{P}}(l^+,l^+,\Delta\tau)$ (in units of
$\Delta\tau a^4/4$ vs. $\Delta \tau$ with
$\alpha,\beta,\gamma,\omega$ non zero as dashed curve and
$\alpha,\beta,\gamma,\omega=0$ is the solid curve, and the
long-dashed curve denotes the case where only $\omega \ne 0$.}}
\label{lpm}
\end{figure}

We next look at the semileptonic decays of Kaons. First we consider the case where both Kaons
decay to $\ell^+$. As discussed in the beginning of the section,
(\ref{obsfnal})
the relevant observable is $\mathcal{O}_{l+}$,
 \ban
 \mathcal{O}_{l+}=\frac{|a|^2}{2}\left(%
\begin{array}{cc}
  1 & 1 \\
  1 & 1 \\
\end{array}%
\right)
 \ean
Recalling that $\epsilon_L=\epsilon_M-\Delta$ and
$\epsilon_S=\epsilon_M+\Delta$ so $\sum
(\epsilon_{SL}+\epsilon_{SL}^{*})=4{\rm Re}\epsilon_M$
we have for the double-decay rate:
 \ba
    \mathcal{P}(l^{\pm} ,\tau_1;l^{\pm},\tau_2)&=&
     \frac{a^4}{8}\left[ \left(1\pm 4{\rm Re}(\epsilon_M)
    \right)  ( e^{-\Gamma_L \tau_2
    -\Gamma_S \tau_1}+
    e^{-\Gamma_L \tau_1-\Gamma_S \tau_2})\right.
    \nonumber\\ && -2(1\pm 4{\rm Re}(\epsilon_M)
    )\cos(\Delta m (\tau_1-\tau_2))e^{-(\overline{\Gamma}+\alpha-\gamma)(\tau_1+\tau_2)}
    \nonumber \\ && -2|\omega|^2\cos(\Delta m
    (\tau_1+\tau_2))e^{-(\overline{\Gamma}+\alpha-\gamma)(\tau_1+\tau_2) }
     +|\omega|^2\left(e^{-\Gamma_S (\tau_1+\tau_2)}
    + e^{-\Gamma_L (\tau_1+\tau_2)}\right)
    \nonumber \\ && + \frac{2\gamma}{\Delta \Gamma}\left(e^{-\Gamma_L (\tau_1+\tau_2)}
    -e^{-\Gamma_S (\tau_1+\tau_2)}\right)
     + \frac{2\alpha}{\Delta m}
    \sin(\Delta m (\tau_1+\tau_2))e^{-(\overline{\Gamma}+\alpha-\gamma)(\tau_1+\tau_2) }
    \nonumber \\ && \pm \left(\frac{4\beta}{|d|}\sin(\Delta m \tau_1 -\phi_{SW})
    +2|\omega|\cos(\Delta m \tau_1-\Omega)\right)
    e^{-(\overline{\Gamma} +\alpha-\gamma)\tau_1}e^{-\Gamma_S\tau_2}
    \nonumber \\ && \pm\left(\frac{4\beta}{|d|}\sin(\Delta m \tau_2 -\phi_{SW})
    -2|\omega|\cos(\Delta m\tau_2-\Omega)\right)
    e^{-(\overline{\Gamma} +\alpha-\gamma)\tau_2}e^{-\Gamma_S\tau_1}
    \nonumber \\ &&  \pm\left(\frac{4\beta}{|d|}\sin(\Delta m \tau_1
    +\phi_{SW})+2|\omega|\cos(\Delta m\tau_1+\Omega)\right)
    e^{-(\overline{\Gamma} +\alpha-\gamma)\tau_1}e^{-\Gamma_L\tau_2}
    \nonumber \\ && \left. \pm \left(\frac{4\beta}{|d|}\sin(\Delta m \tau_2 +\phi_{SW})
    -2|\omega|\cos(\Delta m\tau_2+\Omega)\right)
    e^{-(\overline{\Gamma} +\alpha-\gamma)\tau_2}e^{-\Gamma_L\tau_1}
    \right]\nonumber \\ &&
    \label{llpm}
 \ea
and the integrated quantity reads:
 \ba
    \overline{\mathcal{P}}(&l^{\pm}& ;l^{\pm},\Delta\tau)=
    \frac{a^4}{16}\left[2\left(1\pm 4{\rm Re}\epsilon_M
    \right)  \frac{ e^{-\Gamma_L \Delta\tau }+
    e^{-\Gamma_S \Delta \tau}}{\Gamma_S+\Gamma_L}\right.
 \nonumber \\ &&-(1\pm 4{\rm Re}\epsilon_M
    )2\cos(\Delta m \Delta\tau)
    \frac{e^{-(\overline{\Gamma}+\alpha-\gamma)\Delta\tau}}
    {(\overline{\Gamma}+\alpha-\gamma)}
   \nonumber \\ && -2|\omega|^2e^{-\Delta\tau(\overline{\Gamma}+\alpha-\gamma)}
    \frac{(\overline{\Gamma}+\alpha-\gamma)\cos(\Delta m\Delta\tau)
    -\Delta m\sin(\Delta m\Delta\tau)}{(\overline{\Gamma}+\alpha-\gamma)^2
    +\Delta m^2}
   \nonumber \\ &&+|\omega|^2\left(\frac{e^{-\Gamma_S
    \Delta\tau}}{\Gamma_S}
    + \frac{e^{-\Gamma_L \Delta \tau}}{\Gamma_L}\right)
    + \frac{2\gamma}{\Delta \Gamma}\left(
    \frac{e^{-\Gamma_L \Delta \tau}}{\Gamma_L}
    -\frac{e^{-\Gamma_S
    \Delta\tau}}{\Gamma_S}\right)
   \nonumber \\ &&
     + \frac{2\alpha}{\Delta
     m}e^{-(\overline{\Gamma}+\alpha-\gamma)\Delta \tau}
    \frac{((\overline{\Gamma}+\alpha-\gamma)\sin(\Delta m \Delta\tau)
    +\Delta m\cos(\Delta m\Delta\tau))}{(\overline{\Gamma}+\alpha-\gamma)^2
    +\Delta m^2}
   \nonumber \\ && \pm \frac{4}
    {\Delta m^2+(\overline{\Gamma}+\alpha-\gamma)^2
    +2(\overline{\Gamma}+\alpha-\gamma)\Gamma_S+\Gamma_S^2}
   \nonumber \\ && \times \left(\frac{2\beta}{|d|}e^{-\Delta\tau\Gamma_S}(\Delta m\cos(\phi_{SW})-
    (\overline{\Gamma}+\alpha-\gamma+\Gamma_S)\sin(\phi_{SW})))\right.
    \nonumber \\ && +|\omega| e^{-\Delta\tau\Gamma_S}(\Delta
    m\sin(\Omega)+(\overline{\Gamma}+\alpha-\gamma+\Gamma_S)\cos(\Omega))
    \nonumber \\ &&  + \frac{2\beta}{|d|}e^{-(\overline{\Gamma}+\alpha-\gamma)\Delta\tau}
    (\Delta m\cos(\Delta m\Delta\tau-\phi_{SW})+
    (\overline{\Gamma}+\alpha-\gamma+\Gamma_S)\sin(\Delta m\Delta\tau-\phi_{SW}))
    \nonumber \\ && \left.+|\omega| e^{-(\overline{\Gamma}+\alpha-\gamma)\Delta\tau}(-\Delta
    m\sin(\Delta m\Delta\tau-\Omega)+(\overline{\Gamma}+\alpha-\gamma+\Gamma_S)
    \cos(\Delta m\Delta\tau-\Omega))\right)
    \nonumber \\ && \pm \frac{4}
    {\Delta m^2+(\overline{\Gamma}+\alpha-\gamma)^2
    +2(\overline{\Gamma}+\alpha-\gamma)\Gamma_L+\Gamma_L^2}
     \nonumber \\ && \times\left(\frac{2\beta}{|d|}e^{-\Delta\tau\Gamma_L}(\Delta
     m\cos(\phi_{SW})+
    (\overline{\Gamma}+\alpha-\gamma+\Gamma_L)\sin(\phi_{SW}))\right.
    \nonumber \\ && +|\omega| e^{-\Delta\tau\Gamma_L}(-\Delta
    m\sin(\Omega)+(\overline{\Gamma}+\alpha-\gamma+\Gamma_L)\cos(\Omega))
    \nonumber \\ &&  +  \frac{2\beta}{|d|}e^{-(\overline{\Gamma}+\alpha-\gamma)\Delta\tau}
    (\Delta m\cos(\Delta m\Delta\tau-\phi_{SW})+
    (\overline{\Gamma}+\alpha-\gamma+\Gamma_L)\sin(\Delta
    m\Delta\tau-\phi_{SW}))
    \nonumber \\ && \left.
     \left.+|\omega| e^{-(\overline{\Gamma}+\alpha-\gamma)\Delta\tau}(-\Delta
    m\sin(\Delta m\Delta\tau+\Omega)+(\overline{\Gamma}+\alpha-\gamma+\Gamma_L)
    \cos(\Delta m\Delta\tau+\Omega))\right)\right]\nonumber \\ &&
    \label{llpmint}
\ea
We plot the corresponding integrated double-decay rate for $\ell^+\ell^+$
in Fig. \ref{lpm}.

Next we consider the case in which one of the Kaons decays to $\ell^+$ and the other
to $\ell^-$. On noting that the relevant observable corresponding to the
$\ell^-$  case is associated with the operator $\mathcal{O}_{l-}$ of
(\ref{obsfnal}):
 \ban
 \mathcal{O}_{l-}=\frac{|a|^2}{2}\left(%
\begin{array}{cc}
  1 & -1 \\
  -1 & 1 \\
\end{array}%
\right)
 \ean
the relevant double-decay rate reads:
 \ba
    \mathcal{P}(l^{+} ,\tau_1;l^{-},\tau_2)&=&
     \frac{a^4}{8} \left[(1+4{\rm Re}(\Delta -\beta/d))
     e^{-\Gamma_L \tau_2
    -\Gamma_S \tau_1}\right.
    + (1+4{\rm Re}(\beta/d-\Delta ))
    e^{-\Gamma_L \tau_1-\Gamma_S \tau_2}
    \nonumber \\ && +2\left(\cos(\Delta m\Delta\tau )+4{\rm Im}(\Delta
    +\beta/d)\sin(\Delta m\Delta\tau)\right)e^{-(\overline{\Gamma}+\alpha-\gamma)(\tau_1+\tau_2)}
    \nonumber \\ && +\left( 2|\omega|^2 \cos(\Delta m (\tau_1+\tau_2))
    - \frac{2\alpha}{\Delta m}\sin(\Delta m
    (\tau_1+\tau_2))\right)e^{-(\overline{\Gamma}+\alpha-\gamma)(\tau_1+\tau_2) }
    \nonumber \\ && +\left( \left(|\omega|^2-\frac{2\gamma}{\Delta \Gamma}\right)
    e^{-\Gamma_S (\tau_1+\tau_2)}
    +   \left(|\omega|^2+\frac{2\gamma}{\Delta \Gamma}\right)
     e^{-\Gamma_L (\tau_1+\tau_2)}\right)
     \nonumber \\ && +\left(\frac{4\beta}{|d|}\sin(\Delta m \tau_1 -\phi_{SW})
    +2|\omega|\cos(\Delta m \tau_1-\Omega)\right)
    e^{-(\overline{\Gamma} +\alpha-\gamma)\tau_1}e^{-\Gamma_S\tau_2}
    \nonumber \\ && -\left(\frac{4\beta}{|d|}\sin(\Delta m \tau_2 -\phi_{SW})
    -2|\omega|\cos(\Delta m\tau_2-\Omega)\right)
    e^{-(\overline{\Gamma} +\alpha-\gamma)\tau_2}e^{-\Gamma_S\tau_1}
    \nonumber \\ &&  +\left(\frac{4\beta}{|d|}\sin(\Delta m \tau_1
    +\phi_{SW})+2|\omega|\cos(\Delta m\tau_1+\Omega)\right)
    e^{-(\overline{\Gamma} +\alpha-\gamma)\tau_1}e^{-\Gamma_L\tau_2}
    \nonumber \\ && \left. -\left(\frac{4\beta}{|d|}\sin(\Delta m \tau_2 +\phi_{SW})
    -2|\omega|\cos(\Delta m\tau_2+\Omega)\right)
    e^{-(\overline{\Gamma} +\alpha-\gamma)\tau_2}e^{-\Gamma_L\tau_1}
    \right]\nonumber \\ &&
    \label{lpleftlmright}
 \ea
and the integrated over time distribution is:
  \ba
    &\overline{\mathcal{P}}&(l^{+} ;l^{-},\Delta\tau)=
     \frac{a^4}{16} \left[(1+4{\rm Re}(\Delta -\beta/d))
     \frac{2e^{-\Gamma_L \Delta\tau }}{\Gamma_S+\Gamma_L}\right.
    + (1+4{\rm Re}(\beta/d-\Delta ))
    \frac{2e^{-\Gamma_S \Delta\tau}}{\Gamma_S+\Gamma_L}
   \nonumber \\ && + 2\left(\cos(\Delta m\Delta\tau )+4{\rm Im}(\Delta
    +\beta/d)\sin(\Delta
    m\Delta\tau)\right) \frac{e^{-(\overline{\Gamma}+\alpha-\gamma)\Delta\tau}}
    {(\overline{\Gamma}+\alpha-\gamma)}
   \nonumber \\ && + \frac{e^{-(\overline{\Gamma}+\alpha-\gamma)\Delta\tau}}
    {(\overline{\Gamma}+\alpha-\gamma)+\Delta m^2}(2|\omega|^2((\overline{\Gamma}+\alpha-\gamma)
    \cos(\Delta m\Delta\tau)-\Delta m\sin(\Delta m \Delta\tau))
   \nonumber \\ && - \frac{2\alpha}{\Delta m}(\Delta m\cos(\Delta m\Delta\tau)
    +(\overline{\Gamma}+\alpha-\gamma)\sin(\Delta m\Delta\tau)))
   \nonumber\\ && + \left( \left(|\omega|^2-\frac{2\gamma}{\Delta \Gamma}\right)\frac{e^{-\Gamma_S
    \Delta \tau}}{\Gamma_S}
    +   \left(|\omega|^2+\frac{2\gamma}{\Delta \Gamma}\right)
    \frac{e^{-\Gamma_L \Delta\tau}}{\Gamma_L}\right)
   \nonumber \\ && + \frac{4}
    {\Delta m^2+(\overline{\Gamma}+\alpha-\gamma)^2
    +2(\overline{\Gamma}+\alpha-\gamma)\Gamma_S+\Gamma_S^2}
   \nonumber \\ && \times \left(\frac{2\beta}{|d|}e^{-\Delta\tau\Gamma_S}(\Delta m\cos(\phi_{SW})-
    (\overline{\Gamma}+\alpha-\gamma+\Gamma_S)\sin(\phi_{SW}))\right.
   \nonumber \\ && +|\omega| e^{-\Delta\tau\Gamma_S}(\Delta
    m\sin(\Omega)+(\overline{\Gamma}+\alpha-\gamma+\Gamma_S)\cos(\Omega))
   \nonumber \\ &&  - \frac{2\beta}{|d|}e^{-(\overline{\Gamma}+\alpha-\gamma)\Delta\tau}
    (\Delta m\cos(\Delta m\Delta\tau-\phi_{SW})+
    (\overline{\Gamma}+\alpha-\gamma+\Gamma_S)\sin(\Delta m\Delta\tau-\phi_{SW}))
   \nonumber \\ && \left.-|\omega| e^{-(\overline{\Gamma}+\alpha-\gamma)\Delta\tau}(-\Delta
    m\sin(\Delta m\Delta\tau-\Omega)+(\overline{\Gamma}+\alpha-\gamma+\Gamma_S)
    \cos(\Delta m\Delta\tau-\Omega))\right)
   \nonumber \\ && + \frac{4}
    {\Delta m^2+(\overline{\Gamma}+\alpha-\gamma)^2
    +2(\overline{\Gamma}+\alpha-\gamma)\Gamma_L+\Gamma_L^2}
   \nonumber \\ && \times \left(\frac{2\beta}{|d|}e^{-\Delta\tau\Gamma_L}(\Delta
     m\cos(\phi_{SW})+
    (\overline{\Gamma}+\alpha-\gamma+\Gamma_L)\sin(\phi_{SW}))\right.
   \nonumber \\ && +|\omega| e^{-\Delta\tau\Gamma_L}(-\Delta
    m\sin(\Omega)+(\overline{\Gamma}+\alpha-\gamma+\Gamma_L)\cos(\Omega))
   \nonumber \\ &&   - \frac{2\beta}{|d|}e^{-(\overline{\Gamma}+\alpha-\gamma)\Delta\tau}
    (\Delta m\cos(\Delta m\Delta\tau-\phi_{SW})+
    (\overline{\Gamma}+\alpha-\gamma+\Gamma_L)\sin(\Delta
    m\Delta\tau-\phi_{SW}))
   \nonumber \\ && \left.
     \left.+|\omega| e^{-(\overline{\Gamma}+\alpha-\gamma)\Delta\tau}(-\Delta
    m\sin(\Delta m\Delta\tau+\Omega)+(\overline{\Gamma}+\alpha-\gamma+\Gamma_L)
    \cos(\Delta m\Delta\tau+\Omega))\right)\right]\nonumber \\ &&
    \label{lpleftlmrightint}
 \eea

\begin{figure}[tb]
\includegraphics[width=8cm]{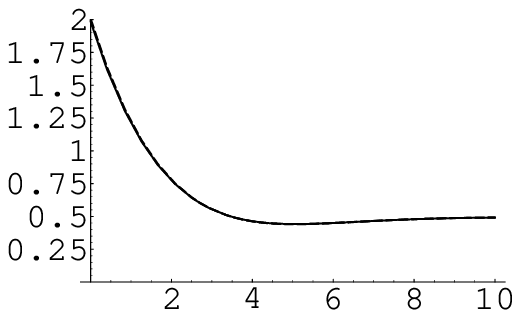}
\includegraphics[width=8cm]{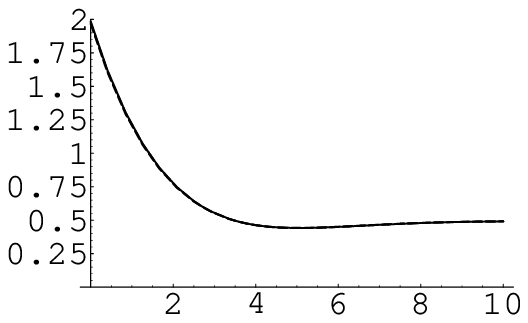}
\includegraphics[width=8cm]{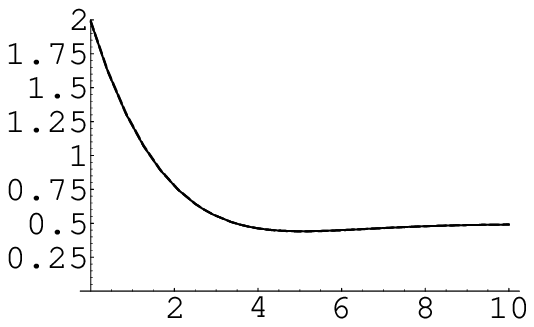}
\includegraphics[width=8cm]{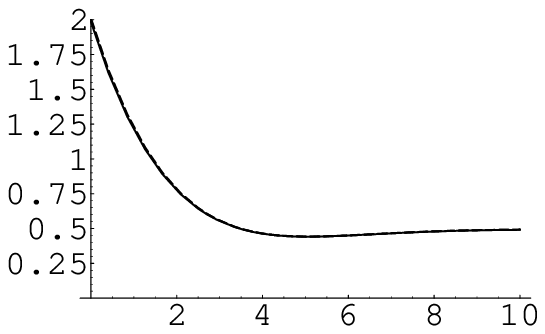}
\caption{{\it Plots of
$\overline{\mathcal{P}}(l^{+};l^{-},\Delta\tau)$, in units of
$\tau_S\frac{|a|^4}{4}$, vs. $\Delta \tau$ with
$\alpha,\beta,\gamma,\omega$ non zero as dashed curve and
$\alpha,\beta,\gamma,\omega=0$ is the solid curve, and the
long-dashed curve denotes the case where only $\omega \ne 0$. }}
\label{lplm}
\end{figure}

\begin{figure}[tb]
\includegraphics[width=8cm]{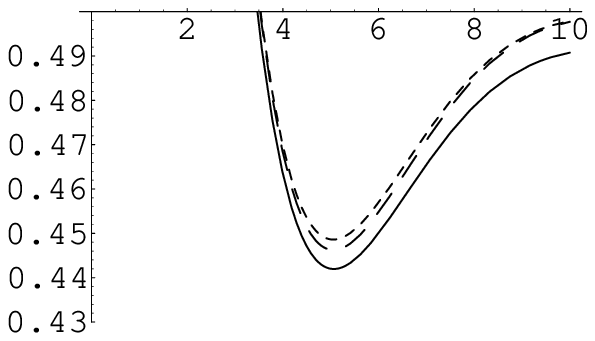}
\includegraphics[width=8cm]{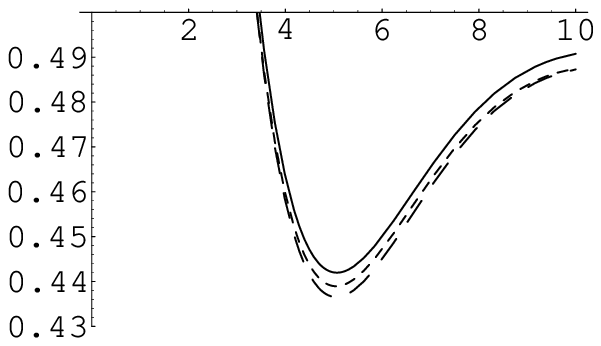}
\includegraphics[width=8cm]{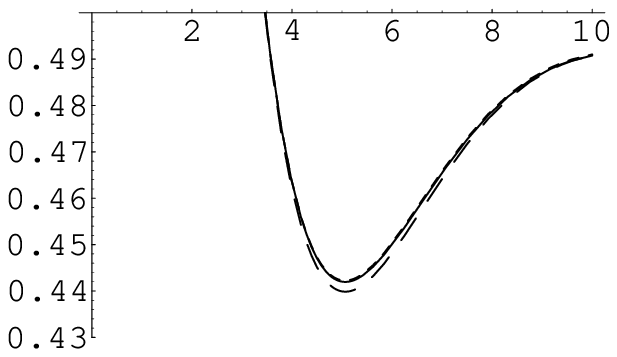}
\includegraphics[width=8cm]{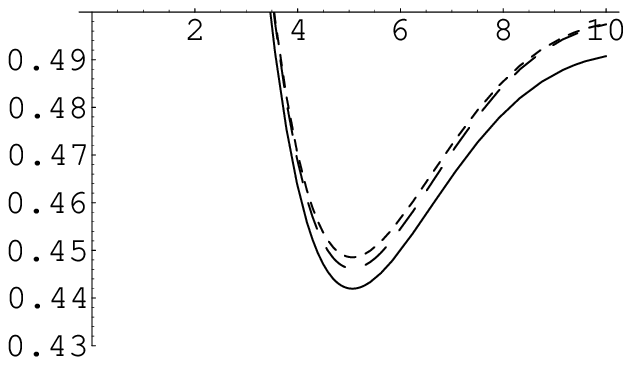}
\caption{{\it As in Fig. \ref{lplm}, but shown in detail to better
illustrate the difference between the various cases. Again, the
dashed curves are with $\alpha,\beta,\gamma,\omega$ all non-zero,
the long-dashed curve has only $\omega \ne 0$, whilst the solid
curve represents $\alpha,\beta,\gamma,\omega=0$.}}
\label{lplmclose}
\end{figure}

\begin{figure}[tb]
\includegraphics[width=8cm]{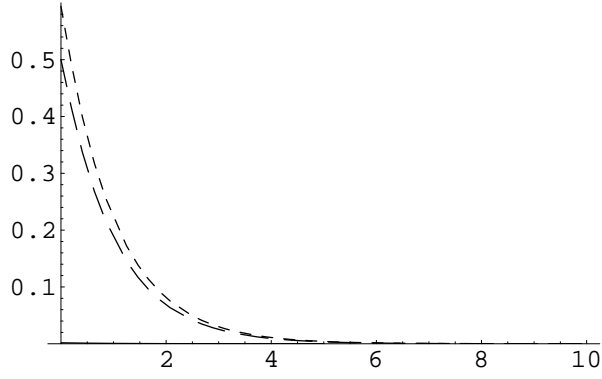}
\caption{{\it Plots of $\overline{\mathcal{P}}(3\pi;
3\pi,\Delta\tau)$, in units of
$\tau_L|\overline{\eta}_{+-}|^2\frac{|X_{3\pi}|^4}{2}$ (for
convenience), vs. $\Delta \tau$ (also in units of $\tau_L$), with
$\alpha,\beta,\gamma,\omega$ non zero as dashed curve and
$\alpha,\beta,\gamma,\omega=0$ is the solid curve, and the
long-dashed curve denotes the case where only $|\omega |=
|\overline{\eta}_{+-}| \ne 0$ (this value is chosen for
concreteness). We see that the CPTV effects are pronounced near
the origin and are easily disentangled from the quantum mechanical
case (which coincides with the horizontal axis).}} \label{3pifig}
\end{figure}

We next give the three pion decay channels using the
relevant observable (\ref{obsfnal}),
\begin{equation}
    \mathcal{O}_{3\pi}=|X_{3\pi}|^2\left(%
\begin{array}{cc}
  |Y_{3\pi}|^2 & Y_{3\pi}^* \\
  Y_{3\pi} & 1 \\
\end{array}%
\right).
\end{equation}
Here we take
 $|Y_{3\pi}|=0$, since we ignore the associated CP and CPT violation
in the decay and obtain:

\begin{equation}
    \mathcal{O}_{3\pi}=|X_{3\pi}|^2\left(%
\begin{array}{cc}
  0 & 0 \\
  0 & 1 \\
\end{array}%
\right).
\end{equation}
This leads to the following expression for the associated
double-time distribution
 \bea
    \mathcal{P}& &(3\pi,\tau_1;3\pi,\tau_2)
    = \frac{|X_{3\pi}|^4}{2}\left[R_S e^{-\Gamma_L \tau_2 -\Gamma_S \tau_1} + R_S
    e^{-\Gamma_L \tau_1-\Gamma_S \tau_2}\right.
    \nonumber\\&&-2|\overline{\eta}_{3\pi}|^2
    \cos(\Delta m (\tau_1-\tau_2))
    e^{-(\overline{\Gamma}+\alpha-\gamma)(\tau_1+\tau_2)}
   +\left(|\omega|^2 + \frac{2\gamma}{\Delta \Gamma}+\frac{4\beta}{|d|}
   |\overline{\eta}_{3\pi}|\frac{\sin\phi_{3\pi}}{\cos\phi_{SW}}\right)
    e^{-\Gamma_L (\tau_1+\tau_2)}
   \nonumber\\&&
     +|\overline{\eta}_{3\pi}|e^{ -\Gamma_L \tau_1
    -(\overline{\Gamma}+\alpha-\gamma)\tau_2}
    \left(-2|\omega|  \cos(\Delta m \tau_2 -\phi_{3\pi}+\Omega)
    +\frac{4\beta}{|d|} \sin(\Delta m\tau_2-\phi_{3\pi}+\phi_{SW})
     \right)
   \nonumber\\&& \left.
     +|\overline{\eta}_{3\pi}|e^{ -\Gamma_L \tau_2
    -(\overline{\Gamma}+\alpha-\gamma)\tau_1}
    \left(2|\omega|  \cos(\Delta m \tau_1 -\phi_{3\pi}+\Omega)
    +\frac{4\beta}{|d|} \sin(\Delta m\tau_1-\phi_{3\pi}+\phi_{SW})
     \right)\right],\nonumber
     \\&&
 \eea
which integrated over time gives
 \bea
    \overline{\mathcal{P}}(3\pi,\tau_1&;&3\pi,\tau_2)
    = \frac{|X_{3\pi}|^4}{4}\left[\frac{2R_S (e^{-\Gamma_L \Delta\tau} +
    e^{-\Gamma_S \Delta\tau})}{\Gamma_S +\Gamma_L}\right.
    \nonumber\\&&-2|\overline{\eta}_{3\pi}|^2
    \cos(\Delta m \Delta\tau)
    \frac{e^{-(\overline{\Gamma}+\alpha-\gamma)\Delta\tau}}
    {\overline{\Gamma}+\alpha-\gamma}
   +\left(|\omega|^2 + \frac{2\gamma}{\Delta \Gamma}+\frac{4\beta}{|d|}
   |\overline{\eta}_{3\pi}|\frac{\sin\phi_{3\pi}}{\cos\phi_{SW}}\right)
    \frac{e^{-\Gamma_L \Delta\tau}}{\Gamma_L}
   \nonumber\\&&
    +\frac{4|\overline{\eta}_{3\pi}|
    e^{ -(\overline{\Gamma}+\alpha-\gamma)\Delta\tau}}
    {\Delta m^2 +(\overline{\Gamma}+\alpha-\gamma)^2
    +2(\overline{\Gamma}+\alpha-\gamma)\Gamma_L+\Gamma_L^2}
    \nonumber \\&&
     \times
    \left(|\omega|e^{ -(\overline{\Gamma}+\alpha-\gamma)\Delta\tau}
     ( -(\overline{\Gamma}+\alpha-\gamma+\Gamma_L)
    \cos(\Delta m \Delta\tau -\phi_{3\pi}+\Omega)\right.
    \nonumber\\&& \quad+\Delta m
    \sin(\Delta m \Delta\tau -\phi_{3\pi}+\Omega))
   \nonumber \\ && +\frac{2\beta}{|d|}
   e^{ -(\overline{\Gamma}+\alpha-\gamma)\Delta\tau}(
   (\overline{\Gamma}+\alpha-\gamma+\Gamma_L)
   \sin(\Delta m\Delta\tau-\phi_{3\pi}+\phi_{SW})
  \nonumber \\ && \quad\left. +\Delta m\cos(\Delta
  m\Delta\tau-\phi_{3\pi}+\phi_{SW})
     \right)
   \nonumber\\&&
     +|\omega|e^{ -\Gamma_L\Delta\tau}
    \left( (\overline{\Gamma}+\alpha-\gamma+\Gamma_L)
     \cos( \Omega-\phi_{3\pi})-\Delta m
     \sin(\Omega-\phi_{3\pi}))\right.
    \nonumber \\&& \left. \left.+
    \frac{2\beta}{|d|}e^{ -\Gamma_L\Delta\tau}(
    (\overline{\Gamma}+\alpha-\gamma+\Gamma_L)\sin(\phi_{SW}-\phi_{3\pi})
    +\Delta m\cos(\phi_{SW}-\phi_{3\pi}))
     \right)\right]. \nonumber
     \\&&
\label{3pion}
 \eea
We plot this function vs. $\Delta \tau$ in Fig. \ref{3pifig}.
We use units of $\tau_L$ for convenience.
The quantum-mechanical
case with $\omega = \alpha = \beta = \gamma = 0$ coincides with the
horizontal axis.
As we observe from the graph, the CPTV and decoherence effects
are pronounced near $\Delta \tau = 0$. Thus, in principle this is the
cleanest way of bounding such effects, but unfortunately
in practice this is a very difficult channel to measure experimentally.

This completes the analysis of identical-final-states observables.

\subsection{General Final States}

We consider in this subsection
observables for the case where the final states are different.

We first consider the case in which one Kaon decays to
$\pi^+\pi^-$ and the other to $\pi^0\pi^0$.
For the double-decay rate we find
 \bea
    &\mathcal{P}&(\pi^+\pi^-, \tau_1; \pi^0 \pi^0,\tau_2)
     = 2A_0^4
    \left[R_L^{00} e^{-\Gamma_L \tau_2-\Gamma_S \tau_1}
    + R_L^{+-} e^{-\Gamma_L \tau_1-\Gamma_S \tau_2}\right.
    \nonumber\\&&-2|\overline{\eta}_{+-}||\overline{\eta}_{00}|
    \cos(\Delta m(\tau_1-\tau_2) +\phi_{+-}-\phi_{00})
    e^{-(\overline{\Gamma}+\alpha-\gamma)(\tau_1+\tau_2)}
   \nonumber \\&&+\left(|\omega|^2- \frac{2\gamma}{\Delta \Gamma}-\frac{2\beta }{|d|}
  \frac{ (|\overline{\eta}_{+-}|\sin{\phi_{+-}}
  + |\overline{\eta}_{00}|\sin{\phi_{00}})}{\cos\phi_{SW}}\right)
     e^{-\Gamma_S (\tau_1+\tau_2)}
    \nonumber\\&& +|\overline{\eta}_{+-}| e^{-\Gamma_S
    \tau_2 -(\overline{\Gamma}+\alpha-\gamma) \tau_1}
    \left(2|\omega| \cos(\Delta m \tau_1 +\phi_{+-}-\Omega)
    +\frac{4\beta}{|d|}\sin(\Delta m\tau_1 +\phi_{+-}-\phi_{SW})\right)
   \nonumber\\&& \left. +|\overline{\eta}_{00}| e^{-\Gamma_S
    \tau_1 -(\overline{\Gamma}+\alpha-\gamma) \tau_2}
    \left(-2|\omega| \cos(\Delta m \tau_2 +\phi_{00}-\Omega)
    +\frac{4\beta}{|d|}\sin(\Delta m\tau_2
    +\phi_{00}-\phi_{SW})\right)\right].
    \nonumber \\&&
 \eea
 The integrated-over-time distribution is given by
 \bea
    &\overline{\mathcal{P}}&(\pi^+\pi^-; \pi^0
    \pi^0;\Delta\tau)=2A_0^4
    \left[R_L^{00}\frac{ e^{-\Gamma_L \Delta\tau}}{\Gamma_L+\Gamma_S}
    + R_L^{+-} \frac{e^{-\Gamma_S
    \Delta\tau}}{\Gamma_L+\Gamma_S}\right.
    \nonumber\\&&-|\overline{\eta}_{+-}||\overline{\eta}_{00}|
    \cos(\Delta m\Delta\tau +\phi_{+-}-\phi_{00})
    \frac{e^{-(\overline{\Gamma}+\alpha-\gamma)\Delta\tau}}
    {\overline{\Gamma}+\alpha-\gamma}
   \nonumber \\ && +\left(|\omega|^2- \frac{2\gamma}{\Delta \Gamma}-\frac{2\beta }{|d|}
  \frac{ (|\overline{\eta}_{+-}|\sin{\phi_{+-}}
  + |\overline{\eta}_{00}|\sin{\phi_{00}})}{\cos\phi_{SW}}\right)
     \frac{e^{-\Gamma_S \Delta \tau}}{2\Gamma_S}
    \nonumber\\ && +\frac{2}{\Delta m^2 +(\overline{\Gamma}+\alpha-\gamma)^2
    +2(\overline{\Gamma}+\alpha-\gamma)\Gamma_S+\Gamma_S^2}
   \nonumber\\&& \times\left[|\overline{\eta}_{+-}| e^{-\Gamma_S
    \Delta\tau }
    \left( |\omega|( (\overline{\Gamma}+\alpha-\gamma+\Gamma_S)
    \cos(\phi_{+-}-\Omega)-\Delta
    m\sin(\phi_{+-}-\Omega))\right.\right.
   \nonumber \\&& \left. +\frac{2\beta}{|d|}(
   (\overline{\Gamma}+\alpha-\gamma+\Gamma_S)
   \sin(\phi_{+-}-\phi_{SW})+\Delta m\cos(\phi_{+-}-\phi_{SW}))\right)
   \nonumber\\&& +|\overline{\eta}_{00}|
   e^{-(\overline{\Gamma}+\alpha-\gamma) \Delta\tau}
    \left(|\omega|( -(\overline{\Gamma}+\alpha-\gamma+\Gamma_S)
    \cos(\Delta m \Delta\tau +\phi_{00}-\Omega)
    +\Delta m\sin(\Delta m \Delta\tau +\phi_{00}-\Omega))\right.
    \nonumber \\ && \left.\left. \left. +\frac{2\beta}{|d|}
    (\overline{\Gamma}+\alpha-\gamma+\Gamma_S)(\sin(\Delta m\Delta\tau
    +\phi_{00}-\phi_{SW})+\Delta m\cos(\Delta m\Delta\tau
    +\phi_{00}-\phi_{SW}))\right)\right]\right].
    \nonumber \\&&
\label{pipmpi0}
\eea

\begin{figure}[tb]
\includegraphics[width=8cm]{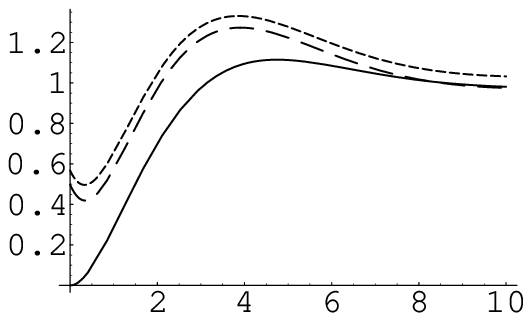}
\includegraphics[width=8cm]{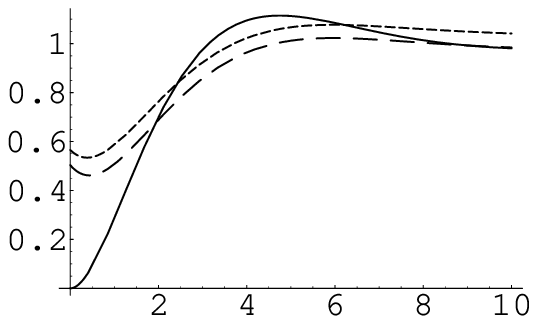}
\includegraphics[width=8cm]{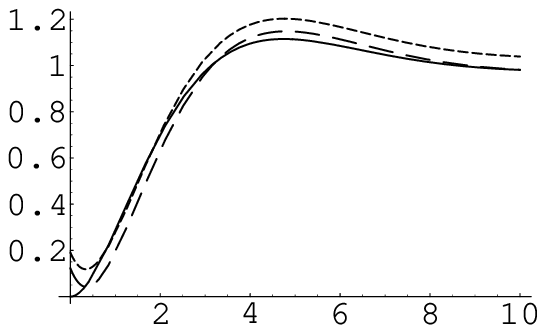}
\includegraphics[width=8cm]{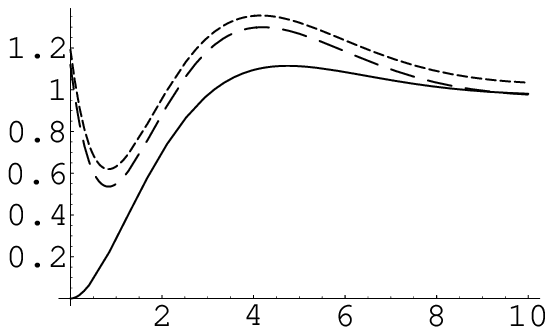}
\caption{{\it Plots of
$\overline{\mathcal{P}}(\pi^+\pi^-;\pi^0\pi^0;\Delta\tau)$, in
units of $\tau_S |\overline{\eta}_{+-}|^2
\frac{|X_{+-}|^2|X_{00}|^2}{2}$, vs. $\Delta \tau$ with
$\alpha,\beta,\gamma,\omega$ non zero as dashed curve and
$\alpha,\beta,\gamma,\omega=0$ is the solid curve, and the
long-dashed curve denotes the case where only $\omega \ne 0$. }}
\end{figure}

Finally we consider
the case in which one of the final states consists of two pions,
and the other is the result of a kaon semileptonic decay. The relevant
double-decay rate is given by:

 \bea
    \mathcal{P}&&(\pi^{\pm} ,\tau_1;l^{\pm},\tau_2)=
    \frac{A_0^2a^2}{4}\left[ (1\pm \delta_L+\frac{\gamma}{\Delta\Gamma})
     e^{-\Gamma_L \tau_2 -\Gamma_S \tau_1}\right.
    + R_L
    e^{-\Gamma_L \tau_1-\Gamma_S \tau_2}
    \nonumber \\ && \mp 2|\overline{\eta}_{+-}|\cos(\Delta m(\tau_1-\tau_2)+\phi_{+-})
    e^{-(\overline{\Gamma}+\alpha-\gamma)(\tau_1+\tau_2)}
     +\left(|\omega|^2 - \frac{2\gamma}{\Delta \Gamma}-\frac{4\beta|\overline{\eta}_{+-}|}{|d|}
     \frac{\sin\phi_{+-}}{\cos\phi_{SW}}\right)
     e^{-\Gamma_S (\tau_1+\tau_2)}
    \nonumber \\ && \left.\pm \left( \frac{4\beta}{|d|}\sin(\Delta m \tau_2-\phi_{SW})
    -2|\omega| \cos(\Delta m \tau_2-\Omega)  \right) e^{ -\Gamma_S
    \tau_1 -(\overline{\Gamma}+\alpha-\gamma)\tau_2} \right]\nonumber \\ &&
    \label{twopionl}
 \eea
where we used the definition $\delta_L=2{\rm Re}\epsilon_L^+$.

The integrated double-decay rate reads:
 \bea
    &\overline{\mathcal{P}}&(\pi^{\pm};l^{\pm},\Delta\tau)=
     \frac{A_0^2a^2}{4}\left[ (1\pm \delta_L+\frac{\gamma}{\Delta\Gamma})
    \frac{e^{-\Gamma_L \Delta\tau }}{\Gamma_L+\Gamma_S}\right.
    + R_L \frac{e^{-\Gamma_S \Delta\tau}}{\Gamma_L+\Gamma_S}
   \nonumber \\ && \mp |\overline{\eta}_{+-}|\cos(\Delta m\Delta\tau+\phi_{+-})
    \frac{e^{-(\overline{\Gamma}+\alpha-\gamma)\Delta\tau}}{(\overline{\Gamma}+\alpha-\gamma)}
     +\left(|\omega|^2 - \frac{2\gamma}{\Delta \Gamma}
     -\frac{4\beta|\overline{\eta}_{+-}|}{|d|}
     \frac{\sin\phi_{+-}}{\cos\phi_{SW}}\right)
     \frac{e^{-\Gamma_S \Delta\tau}}{2\Gamma_S}
   \nonumber \\ && \pm \frac{2}
    {\Delta m^2+(\overline{\Gamma}+\alpha-\gamma)^2
    +2(\overline{\Gamma}+\alpha-\gamma)\Gamma_S+\Gamma_S^2}
  \nonumber \\ &&  + \left(\frac{2\beta}{|d|}e^{-(\overline{\Gamma}+\alpha-\gamma)\Delta\tau}
    (\Delta m\cos(\Delta m\Delta\tau-\phi_{SW})-
    (\overline{\Gamma}+\alpha-\gamma+\Gamma_S)\sin(\Delta
    m\Delta\tau-\phi_{SW}))\right.
  \nonumber \\ && \left.-|\omega| e^{-(\overline{\Gamma}+\alpha-\gamma)\Delta\tau}(-\Delta
    m\sin(\Delta m\Delta\tau-\Omega)+(\overline{\Gamma}+\alpha-\gamma+\Gamma_S)
    \cos(\Delta m\Delta\tau-\Omega))\right)\nonumber \\ &&
  \label{twopionlint}
 \eea

\begin{figure}[tb]
\includegraphics[width=8cm]{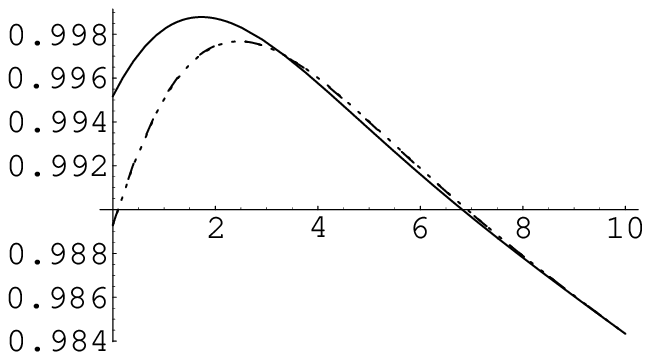}
\includegraphics[width=8cm]{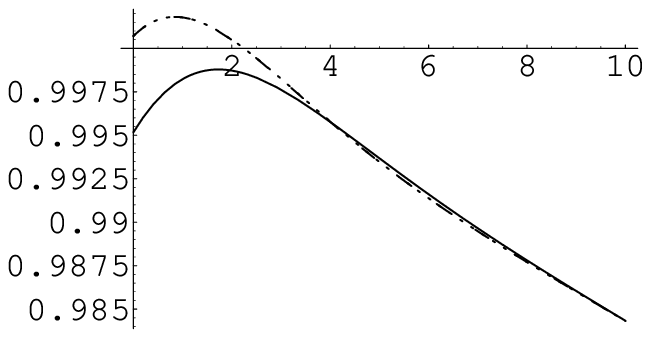}
\includegraphics[width=8cm]{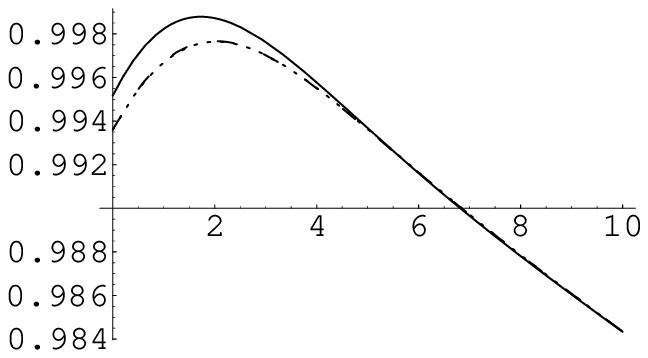}
\includegraphics[width=8cm]{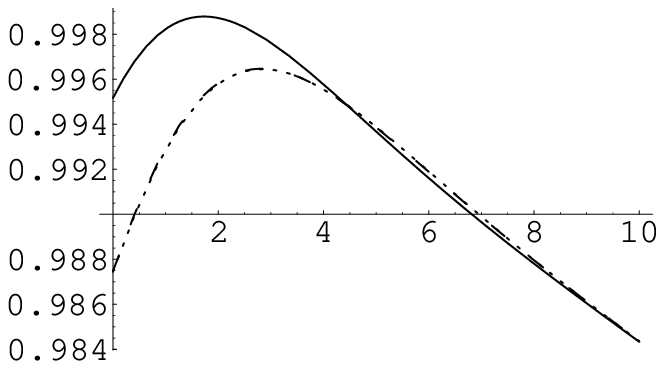}
\caption{{\it Plots of
$\overline{\mathcal{P}}(\pi^{\pm};l^{\pm},\Delta\tau)$, in units
of $\tau_S \frac{|X_{+-}|^2|a|^2}{8}$, vs. $\Delta \tau$ with
$\alpha,\beta,\gamma,\omega$ non zero as dashed curve and
$\alpha,\beta,\gamma,\omega=0$ is the solid curve, and the
long-dashed curve denotes the case where only $\omega \ne 0$.
Notice that in the scale chosen, the dashed and long-dashed curves
overlap completely.}}
\end{figure}

This completes our analysis of observables in a $\phi$ factory that
can be used as sensitive probes for tests of possible
CPTV and quantum decoherence,
to leading order in the small parameters parametrizing
the effects.

\section{Discussion and Conclusions}

In this work, we have embarked into a combined treatment
of decoherent and intrinsic CPTV effects in a $\phi$ factory.
By studying a variety of observables, involving identical as well as
general final states, we have derived analytical expressions
for double-decay rates and their integrated counterparts
(over time $\tau_1 + \tau_2$,
keeping $\Delta \tau = \tau_1 - \tau_2$ fixed) , which we plotted as functions
of $\Delta \tau$. Our analysis included $\omega$, as well as decoherence
$\alpha,\beta,\gamma$ effects. We presented the results to leading order in
these small parameters.

Although the pertinent formulae are algebraically rather complex,
nevertheless
from our general study it became evident that one may disentangle
the $\omega$ from the decoherent evolution effects by looking
simultaneously at different
regimes of $\Delta \tau$, namely, (i) $\Delta \tau = 0$, (ii)
intermediate values of $\Delta \tau$ (as compared with $\tau_S$, which sets a convenient
characteristic time scale in the problem),
with emphasis on interference terms with sinusoidal time dependence,
and (iii) rather large values of
$\Delta \tau $, with emphasis on the exponential damping
of the relevant quantities, which is affected (slowed down)
by the presence of decoherence terms $\alpha - \gamma$, in an
$\omega$-independent way.

In addition to these results in a $\phi$ factory, one can obtain
valuable information on the decoherence parameters $\alpha,\beta$,and $\gamma$
by looking at experiments involving single neutral Kaon beams,
such as CPLEAR~\cite{cplear}, which is achieved
through flavour tagging of the neutral Kaon by means of
the electric charge of the pion on the complementary side.
As discussed in \cite{lopez}, the
decoherence parameters $\alpha,\beta,\gamma$ can be separately
disentangled
by combined
studies of several kaon asymmetries, which are again time-dependent
functions. Notice though that, if complete positivity of entangled state
density matrices is imposed~\cite{benatti},
then only one decoherence parameter $\gamma > 0$
survives in the model of \cite{ehns}, which facilitates
the situation enormously.

We would like at this point to make a 
clarification concerning the important differences 
between proposed experiments 
here, using entangled Kaon states, and some existing 
proposals, notably by the CPLEAR collaboration~\cite{cplearepr},
to measure EPR correlations. 
The reader should notice that the 
accuracy for measuring the appropriate observables 
in \cite{cplearepr} appears to be of order  
$10^{-1}$; however, this accuracy pertains only 
to an asymmetry built 
from ${\overline p}p \to K^0{\overline K}^0$ in 
CPLEAR, looking for the dilepton 
decay channel, between ``like'' ($K^0 K^0$ or ${\overline K}^0 {\overline K}^0$)
and ``unlike'' ($K^0 {\overline K}^0$ or ${\overline K}^0 K^0$)
decays after time evolution. 
As shown in our paper, the most interesting 
observable for the $\omega$-effect comes from 
identical decay channels, 
i.e., in this case the ``intra'' asymmetry between 
$K^0 K^0$ and ${\overline K}^0 {\overline K}^0$, which was not discussed in 
ref. \cite{cplearepr}. 
Furthermore, our observation~\cite{bmp} that the channel ($\pi^+\pi^-$)
($\pi^+\pi^-$) is automatically enhanced by three orders of magnitude, 
due to the relative $(\omega/\eta_{+-})$ amplitude
between the ``wrong'' and the ``right'' symmetry states suggests
that even a $10^{-1}$ experimental effect would represent 
a $10^{-4}$ effect in $\omega$. 
In fact, the expectation for the channel $e^+e^- \to K^0{\overline K}^0$ 
in an upgraded $DA\Phi NE$ is definitely much better, 
anticipating an experimental effect of the order of $10^{-4}-10^{-5}$.

In our analysis we saturated the bounds for the decoherence parameters
$\alpha,\beta,\gamma$ obtained from CPLEAR~\cite{cplear}, to disentangle
the $\omega$ effect by looking at decoherent evolution of observables
in a $\phi$ factory. As mentioned above, 
the most sensitive probe appears at first sight to
be the two-charged-pion channels, as a result of enhancement factors
of $\overline{\eta}_{+-}$. In fact the clearest test of deviation from
quantum mechanics is to look at the behaviour of
the pertinent time-integrated decay rate (\ref{twopionint})
near the $\Delta \tau = 0$
regime, where
the novel CPTV effects would yield a non-zero result for the
integrated asymmetry.

However, in  the presence of decoherence  there is a  reduction of the
value of the asymmetry as compared  to the pure $\omega \ne 0$ unitary
case of \cite{bmp}, by shifting  the value of the asymmetry at $\Delta
\tau = 0 $ from $|\omega^2|$ to $|\omega^2|-2\gamma/\Delta \Gamma$ (in
appropriate  units). This  reduction may  be significant  in  the case
where      the      $|\omega|$      effects     are      of      order
$\sqrt{\gamma/\Delta\Gamma}$.     Of    course,    for   a    reliable
order-of-magnitude  estimate of the  $\omega$-parameter, one  needs to
resort to detailed microscopic models of QG space-time foam, a task we
hope to undertake in the  near future.  The fact that the interference
terms  for $\Delta \tau  \ne 0$  of the  asymmetry (\ref{twopionint}),
which  depend  sinusoidally  on  $\Delta \tau$,  are  proportional  to
$|\omega |$,  and they do not  depend (apart from  damping factors) on
decoherence parameters,  allows in principle for  a disentanglement of
$\omega$ from  decoherence ($\gamma$, ...) effects.   Notice also that
another  clear (in  principle)  probe  of such  effects  would be  the
time-integrated  observable (\ref{3pion}), associated  with three-pion
decay channels,  near $\Delta \tau = 0$.   Unfortunately, however, the
current experimental sensitivity for this observable is low.

Before closing we would like to make a few comments on the contamination
of the actual observable for the measurement of $\epsilon '$
by decoherence~\cite{peskin} and intrinsic CPTV effects.
The
traditional observable for the measurement of
$\epsilon '$ is the asymmetry based on charged-pion/neutral-pion
(\ref{pipmpi0})-type
observables:
\bea
&& {\cal A}(\pi^+,\pi^-;\pi^0,\pi^0; \delta \tau) =
\frac{\overline{\mathcal{P}}(\pi^+\pi^-; \pi^0\pi^0;\Delta\tau)
- \overline{\mathcal{P}}(\pi^0\pi^0; \pi^+\pi^-;\Delta\tau)}
{\overline{\mathcal{P}}(\pi^+\pi^-; \pi^0\pi^0;\Delta\tau)
+ \overline{\mathcal{P}}(\pi^0\pi^0; \pi^+\pi^-;\Delta\tau)} \nonumber \\
&& = 3{\rm Re}\frac{\epsilon '}{\epsilon} {\cal I}_1 - 3{\rm
Im}\frac{\epsilon '}{\epsilon} {\cal I}_2 \label{eprimeasym} \eea

In particular we find
 \bea
    {\cal I}_1&=& \frac{1}{\mathcal{D}}\left[ e^{-\Gamma_L\Delta\tau}\left( 1+ 2\frac{\beta}
    {|d||\overline{\eta}_{+-}|}\sin(\phi_{SW}-\phi_{+-})
    \right)\right.
    \nonumber \\ && - e^{-\Gamma_S\Delta\tau} \left( 1+2\frac{\beta}
    {|d||\overline{\eta}_{+-}|}(\sin(\phi_{SW}-\phi_{+-})
    -|z|\sin(\phi_{SW}-\phi_{+-}+\phi_z))\right.
    \nonumber \\ && \quad\left.+\frac{|\omega|}
    {|\overline{\eta}_{+-}|}|z|\cos(\Omega-\phi_{+-}+\phi_z)\right)
    \nonumber \\ && +e^{-(\overline{\Gamma}+\alpha-\gamma)\Delta\tau}
    \left(2\frac{\beta}{|d||\overline{\eta}_{+-}|}
    |z|\sin(\Delta m\Delta\tau +\phi_{+-}-\phi_{SW}-\phi_z)\right.
    \nonumber \\ && \quad\left.\left.-\frac{|\omega|}{|\overline{\eta}_{+-}|}|z|
    \cos(\Delta m\Delta\tau
    -\Omega+\phi_{+-}-\phi_z)\right)\right],
    \nonumber \\ &&
    \nonumber \\ &&
 \eea
 \bea
    {\cal I}_2&=& \frac{1}{\mathcal{D}}\left[ e^{-\Gamma_L\Delta\tau}\left(  2\frac{\beta}
    {|d||\overline{\eta}_{+-}|}\cos(\phi_{SW}-\phi_{+-})
    \right)\right.
    \nonumber \\ && - e^{-\Gamma_S\Delta\tau} \left( 2\frac{\beta}
    {|d||\overline{\eta}_{+-}|}(\cos(\phi_{SW}-\phi_{+-})
    -|z|\cos(\phi_{SW}-\phi_{+-}+\phi_z))\right.
    \nonumber \\ && \quad\left.+\frac{|\omega|}
    {|\overline{\eta}_{+-}|}|z|\sin(\Omega-\phi_{+-}+\phi_z)\right)
    \nonumber \\ && +e^{-(\overline{\Gamma}+\alpha-\gamma)\Delta\tau}
    \left(2\sin(\Delta m\Delta\tau)-2\frac{\beta}{|d||\overline{\eta}_{+-}|}
    |z|\cos(\Delta m\Delta\tau +\phi_{+-}-\phi_{SW}-\phi_z)\right.
    \nonumber \\ && \quad\left.\left.-\frac{|\omega|}{|\overline{\eta}_{+-}|}|z|
    \sin(\Delta m\Delta\tau
    +\Omega-\phi_{SW}-\phi_z)\right)\right],
    \nonumber \\ &&
    \nonumber \\ &&
 \eea
and
 \bea
    \mathcal{D}&=& e^{-\Gamma_L\Delta\tau}\left(
    1+ \frac{\gamma}{\Delta\Gamma|\overline{\eta}_{+-}|^2}
    +2\frac{\beta}{|d||\overline{\eta}_{+-}|}\frac{\sin(2\phi_{SW}-\phi_{+-})}
    {\cos(\phi_{SW})}\right)
    \nonumber \\ && +e^{-\Gamma_S\Delta\tau}\left(1+
    \frac{\gamma}{\Delta\Gamma|\overline{\eta}_{+-}|^2}\frac{\Gamma_L}{\Gamma_S}
    +2\frac{\beta}{|d||\overline{\eta}_{+-}|}\left(\frac{\sin(2\phi_{SW}-\phi_{+-})}
    {\cos(\phi_{SW})}-2|z|\sin(\phi_{SW}+\phi_{z}-\phi_{+-})\right)\right.
    \nonumber \\ &&\left. +\frac{|\omega|^2\overline{\Gamma}}
    {|\overline{\eta}_{+-}|^2 \Gamma_S}
    -2\frac{|\omega|}{|\overline{\eta}_{+-}|}|z|\cos(\Omega +\phi_z-\phi_{+-}) \right)
    \nonumber \\ && -e^{-(\overline{\Gamma}+\alpha-\gamma)}\left(
    2\cos(\Delta m \Delta \tau)-4\frac{\beta}{|d||\overline{\eta}_{+-}|}
    |z|\sin(\Delta m\Delta\tau
    +\phi_{+-}-\phi_{SW}-\phi_z)\right.
    \nonumber \\ && \left.+\frac{2|\omega|}{|\overline{\eta}_{+-}|}
    |z|\cos(\Delta m\Delta\tau +\phi_{+-}-\Omega-\phi_{z})\right)
    \nonumber \\ &&
 \eea
where we have defined~\cite{peskin}: $|z|e^{i\phi_z}=\frac{2\overline{\Gamma}}{\Gamma_S + \overline{\Gamma} + i\Delta m}$.

It therefore becomes clear that, although in conventional situations,
where the foam and intrinsic CPTV $\omega$-effects are absent,
the quantities ${\rm Re}(\epsilon '/\epsilon)$ and
${\rm Im}(\epsilon '/\epsilon)$
can be extracted from a measurement of
${\cal A}(\pi^+,\pi^-;\pi^0,\pi^0; \Delta \tau)$
by a simple two-parameter fit, in the presence of the quantum-gravity effects
this is no longer true. The coefficients ${\cal I}_{i},~i=1,2$
are modified in such a case by decoherence~\cite{peskin}
and $\omega$-dependent terms.

It is worthy of pointing out at this stage that
the $\omega$-dependent terms in the expression for the above asymmetry
are always accompanied by factors $e^{-\Gamma_S \Delta \tau}$.
Therefore, in the limit $\Gamma_S \Delta \tau \gg 1$
such terms are suppressed. We remind the reader that, in this limit,
in conventional situations, one has simply that
${\cal A}(\pi^+,\pi^-;\pi^0,\pi^0; \Delta \tau) \to
3{\rm Re}\frac{\epsilon '}{\epsilon}$. In contrast,
in the CPTV foamy situation the same limit
contains, to leading order,
terms proportional to \emph{both} real and imaginary
parts of $\epsilon '/\epsilon$, with coefficients
dependent on the decoherence parameters, $\beta,\gamma$,
but independent of $\omega$.
It is also worthy of noting
the form of ${\cal I}_2$ in this limit:
${\cal I}_2={\cal O}(\beta )$ to leading order
in the small parameters.
In complete positivity models~\cite{benatti}, therefore, which require
$\beta = 0$, such ${\cal I}_2$ terms are absent from the right-hand-side of
the observable (\ref{eprimeasym}), and the (limiting) result
for the asymmetry involves, in such a case, only
the factor $3{\rm Re}\frac{\epsilon '}{\epsilon}
\left( 1 - |{\cal O}(\gamma/\Delta \Gamma |\overline{\eta}_{+-}|^2)|\right)$
to leading order.

The above issues are important to bear in mind in experimental derivations of
the $\epsilon '$ parameter, which in the general case
require disentanglement of the
(possible) decoherence and $\omega$ effects by means of a combined study of
the $\phi$-factory observables
described in this article and in \cite{peskin}.

\section*{Acknowledgments}

This work has been partially supported by the Grant CICYT FPA2002-00612.
N.E.M. wishes to thank the University of Valencia, Department of
Theoretical Physics and IFIC for the hospitality during the final stages
of the collaboration. A.W.-L. also thanks
the University of Valencia, Department of
Theoretical Physics for the hospitality and support during the early
stages of this work.

\section*{Appendix: Complete Formulae with $\epsilon '$ corrections}

In this Appendix we give the $\epsilon '$ corrections to the
pertinent double-decay rates.
The inclusion of such effects is accounted for by the $Y_{+-}$ parts
of the observables for the Kaon decays to two and three pions  in
(\ref{obsfnal}).

For the two-charged-pion decay we have
\ban
    \mathcal{O}_{+-}=|X_{+-}|^2\left(%
\begin{array}{cc}
  1 & Y_{+-} \\
  Y_{+-}^* & |Y_{+-}|^2 \\
\end{array}%
\right)
 \ean
with $X_{+-}\braket{\pi\pi}{K_2}$
$Y_{\pi\pi}=\frac{\braket{\pi\pi}{K_1}}{\braket{\pi\pi}{K_2}}$
and we remind the reader that the quantum mechanical
amplitudes
$\eta_{+-}=\epsilon +\epsilon'$, $\eta_{00}=\epsilon -2\epsilon'$,
should be replaced now by their barred counterparts $\overline{\eta}$
that include decoherent contributions as well. In particular,
above we have used the relations~\cite{peskin}
$R_L=\frac{\gamma}{\Delta\Gamma}+|\overline{\eta}_{+-}|^2
+4\frac{\beta}{\Delta\Gamma}{\rm
Im}[\overline{\eta}_{+-}d/d^*-Y_{+-}]$ and
$|\overline{\eta}_{+-}|e^{i\phi_{+-}}=\epsilon_L^- +Y^{+-}$
(the reader should recall that $\epsilon_L^-=\epsilon_L-\frac{\beta}{d}$).

The relevant double-decay time distribution
reads then:
\bea
    &\mathcal{P}&(\pi^+ \pi^-,\tau_1;\pi^+ \pi^-,\tau_2)=
    \frac{|X_{+-}|^4}{2}\left[ R_L(
    e^{-\Gamma_L \tau_2 -\Gamma_S \tau_1} +
    e^{-\Gamma_L \tau_1-\Gamma_S \tau_2})\right.
    \nonumber \\ && -2|\overline{\eta}_{+-}|^2\cos(\Delta m
    (\tau_1-\tau_2) e^{-(\overline{\Gamma}+\alpha -\gamma)(\tau_1+\tau_2)}
   \nonumber\\ && +\frac{4\beta |\overline{\eta}_{+-}|}{|d|}
    \sin(\Delta m \tau_1 +\phi_{+-}-\phi_{SW})e^{-(\overline{\Gamma}+\alpha
    -\gamma)\tau_1-\Gamma_S\tau_2}
    \nonumber\\ && +\frac{4\beta |\overline{\eta}_{+-}|}{|d|}
    \sin(\Delta m \tau_2 +\phi_{+-}-\phi_{SW})e^{-(\overline{\Gamma}+\alpha -\gamma)
    \tau_2-\Gamma_S\tau_1}
    \nonumber\\ && +2|\omega||\overline{\eta}_{+-}|(\cos(\Delta m\tau_1+\phi_{+-}-\Omega)
    e^{-\Gamma_S(\tau_2)-(\overline{\Gamma}+\alpha-\gamma)\tau_1}
    -\cos(\Delta m\tau_2+\phi_{+-}-\Omega)
    e^{-\Gamma_S\tau_1-(\overline{\Gamma}+\alpha-\gamma)\tau_2})
    \nonumber \\&&\left.+\left(|\omega|^2-\frac{2\gamma}{\Delta \Gamma}
    -8\frac{\beta}{\Delta\Gamma}{\rm Im}[\overline{\eta}_{+-}-Y_{+-}]\right)
    e^{-\Gamma_S (\tau_1+\tau_2)}
    \right]\nonumber \\ &&
    \label{twopionprime}
 \eea

The time Integrated distribution  reads
 \bea
    &\overline{\mathcal{P}}&(\pi^+ \pi^-;\pi^+ \pi^-;\Delta\tau)=
    \frac{|X_{+-}|^4}{2}
    \left. \bigg[ R_L \frac{e^{-\Delta \tau \Gamma_L}+e^{-\Delta \tau
    \Gamma_S}} {\Gamma_L+\Gamma_S}\right.
   \nonumber \\ &&\left. -|\overline{\eta}_{+-}|^2 \cos(\Delta m\Delta
    \tau)\frac{e^{-(\overline{\Gamma}+\alpha -\gamma)\Delta\tau}}
    {(\overline{\Gamma}+\alpha -\gamma)}
    \right.
    \nonumber \\ &&\left. + \frac{1}{\Delta m^2+(\overline{\Gamma}+\alpha-\gamma)^2
    +2(\overline{\Gamma}+\alpha-\gamma)\Gamma_S+\Gamma_S^2}\right.
   \nonumber\\ &&\left. \times \left. \bigg[ \frac{4\beta
    |\overline{\eta}_{+-}|}{|d|}\left((\Delta m\cos(\phi_{+-}-\phi_{SW})
    +(\overline{\Gamma}+\alpha-\gamma+\Gamma_S)\sin(\phi_{+-}-\phi_{SW}))
    e^{-\Delta\tau \Gamma_S}
    \right.\right.\right.
   \nonumber \\ && \left. +e^{-\Delta\tau(\overline{\Gamma}+\alpha-\gamma)}
    (\Delta m\cos(\Delta m\Delta\tau +\phi_{+-}-\phi_{SW})\right.
   \nonumber \\ && \left.\left.+(\overline{\Gamma}+\alpha-\gamma+\Gamma_S)
    \sin(\Delta m\Delta\tau +\phi_{+-}-\phi_{SW}))
    \right)\right.
    \nonumber\\ && \left. +2|\omega||\overline{\eta}_{+-}|e^{-\Delta\tau \Gamma_S}
    ((\overline{\Gamma}+\alpha-\gamma+\Gamma_S)\cos(\phi_{+-}-\Omega)
    -\Delta m \sin(\phi_{+-}-\Omega))\right.
    \nonumber \\ && \left. -2 |\omega||\overline{\eta}_{+-}| e^{(\overline{\Gamma}+\alpha-\gamma)\Delta\tau}
    ((\overline{\Gamma}+\alpha-\gamma+\Gamma_S)\cos(\Delta m\Delta\tau+\phi_{+-}-\Omega )\right.
   \nonumber \\
&& \left. \left.-\Delta m\sin(\Delta
m\Delta\tau+\phi_{+-}-\Omega))+\left(|\omega|^2-\frac{2\gamma}{\Delta
\Gamma}
    -8\frac{\beta}{\Delta\Gamma}{\rm Im}[\overline{\eta}_{+-}-Y_{+-}] \right)
    \frac{ e^{-\Gamma_S \Delta\tau}}{2\Gamma_S}\right. \bigg]\right. \bigg]
\nonumber \\ && \label{twopionintprime} \eea

\begin{figure}[tb]
\includegraphics[width=8cm]{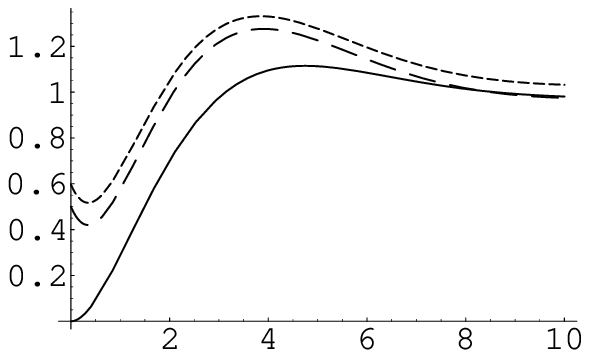}
\includegraphics[width=8cm]{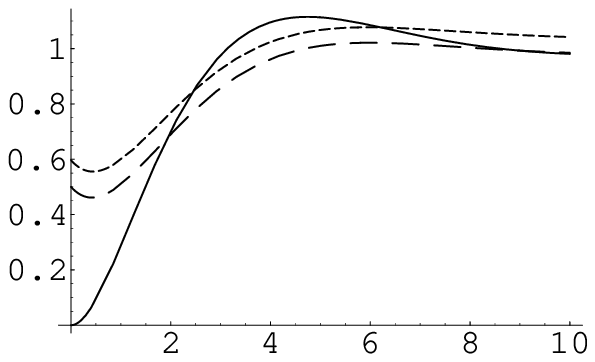}
\includegraphics[width=8cm]{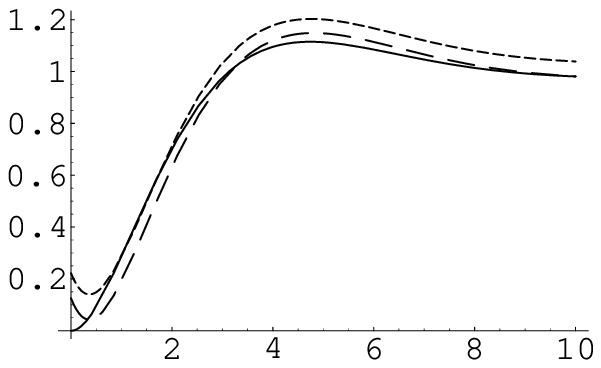}
\includegraphics[width=8cm]{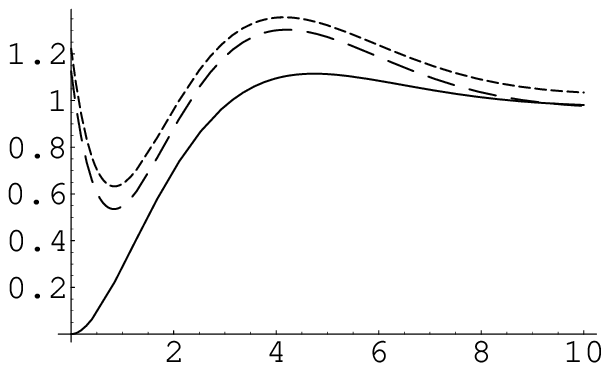}
\caption{{\it Plots of
$\overline{\mathcal{P}}(\pi^{\pm},\pi^{\pm},\Delta\tau)$ vs.
$\Delta \tau$, including $\epsilon '$ corrections. The long-dashed
curve has only $\omega \ne 0$, whilst the solid curve represents
$\alpha,\beta,\gamma,\omega=0$.  The values for $\omega$ and
$\Omega$ are (from left to right, and top to bottom) i)
$|\omega|=|\overline{\eta}_{+-}|$, $\Omega=\phi_{+-}-0.16\pi$, ii)
$|\omega|=|\overline{\eta}_{+-}|$, $\Omega=\phi_{+-}+0.95\pi$,
iii) $|\omega|=0.5|\overline{\eta}_{+-}|$,
$\Omega=\phi_{+-}+0.16\pi$, iv)
$|\omega|=1.5|\overline{\eta}_{+-}|$, $\Omega=\phi_{+-}$.
$\overline{\mathcal{P}}$ is in units of $|\overline{\eta}_{+-}|^2
\tau_s 4 A_0^{4}$ and $\Delta\tau$ is in units of $\tau_S$.}}
\end{figure}

For the neutral-pion decay channel we use the observable:
\ban
    \mathcal{O}_{00}=|X_{00}|^2\left(%
\begin{array}{cc}
  1 & Y_{00} \\
  Y_{00}^* & |Y_{00}|^2 \\
\end{array}%
\right)
 \ean
and the associated double-decay rate involving
charged- and neutral-pion decay channels has the form:
 \bea
    &\mathcal{P}&(\pi^+\pi^-, \tau_1; \pi^0 \pi^0,\tau_2)
     = \frac{|X_{+-}|^2|X_{00}|^2}{2}
    \left[R_L^{00} e^{-\Gamma_L \tau_2-\Gamma_S \tau_1}
    + R_L^{+-} e^{-\Gamma_L \tau_1-\Gamma_S \tau_2}\right.
    \nonumber\\&&-2|\overline{\eta}_{+-}||\overline{\eta}_{00}|
    \cos(\Delta m(\tau_1-\tau_2) +\phi_{+-}-\phi_{00})
    e^{-(\overline{\Gamma}+\alpha-\gamma)(\tau_1+\tau_2)}
    \nonumber\\&& +|\overline{\eta}_{+-}| e^{-\Gamma_S
    \tau_2 -(\overline{\Gamma}+\alpha-\gamma) \tau_1}
    \left(2|\omega| \cos(\Delta m \tau_1 +\phi_{+-}-\Omega)
    +\frac{4\beta}{|d|}\sin(\Delta m\tau_1 +\phi_{+-}-\phi_{SW})\right)
   \nonumber\\&&  +|\overline{\eta}_{00}| e^{-\Gamma_S
    \tau_1 -(\overline{\Gamma}+\alpha-\gamma) \tau_2}
    \left(-2|\omega| \cos(\Delta m \tau_2 +\phi_{00}-\Omega)
    +\frac{4\beta}{|d|}\sin(\Delta m\tau_2
    +\phi_{00}-\phi_{SW})\right)
    \nonumber \\&& \left. +\left(|\omega|^2
    - \frac{2\gamma}{\Delta \Gamma}-4\frac{\beta}{\Delta\Gamma}
    {\rm Im} [\overline{\eta}_{+-}-Y_{+-}+\overline{\eta}_{00}-Y_{00}]\right)
     e^{-\Gamma_S (\tau_1+\tau_2)}\right].
    \nonumber \\&&
 \eea

The double-decay time distribution involving charged-pion and dilepton
channels is given by:
 \bea
    \mathcal{P}(\pi^{\pm} ,\tau_1&;&l^{\pm},\tau_2)=
    \frac{|X_{+-}|^2|a|^2}{4}\left[ (1\pm \delta_L+\frac{\gamma}{\Delta\Gamma})
     e^{-\Gamma_L \tau_2 -\Gamma_S \tau_1}\right.
    + R_L
    e^{-\Gamma_L \tau_1-\Gamma_S \tau_2}
    \nonumber \\ && \mp 2|\overline{\eta}_{+-}|\cos(\Delta m(\tau_1-\tau_2)+\phi_{+-})
    e^{-(\overline{\Gamma}+\alpha-\gamma)(\tau_1+\tau_2)}
    \nonumber \\ && \pm \left( \frac{4\beta}{|d|}\sin(\Delta m \tau_2-\phi_{SW})
    -2|\omega| \cos(\Delta m \tau_2-\Omega)  \right) e^{ -\Gamma_S
    \tau_1 -(\overline{\Gamma}+\alpha-\gamma)\tau_2}
    \nonumber \\ && \left.+\left(|\omega|^2 - \frac{2\gamma}{\Delta \Gamma}
    -\frac{8\beta}{\Delta\Gamma}{\rm Im} [\overline{\eta}_{+-}]\right)
     e^{-\Gamma_S (\tau_1+\tau_2)} \right]\nonumber \\ &&
    \label{twopionlprime}
 \eea

Finally for the three-pion decay channel the relevant observable
is:
\ban
    \mathcal{O}_{3\pi}=|X_{3\pi}|^2\left(%
\begin{array}{cc}
  |Y_{3\pi}|^2 & Y_{3\pi}^* \\
  Y_{3\pi} & 1 \\
\end{array}%
\right)
 \ean
with $\braket{3\pi}{K_2}$
$Y_{3\pi}=\frac{\braket{3\pi}{K_1}}{\braket{3\pi}{K_2}}$

In the following we use
$R_{S}=-\frac{\gamma}{\Delta\Gamma}+|\overline{\eta}_{3\pi}|^2
-4\frac{\beta}{\Delta\Gamma}{\rm Im}[\overline{\eta}_{3\pi}d/d^*
-Y_{3\pi}]$ where
$|\overline{\eta}_{3\pi}|e^{i\phi_{3\pi}}=\epsilon_S^+ +Y_{3\pi}$.

The pertinent three-pion double-decay time distribution is:
 \bea
    \mathcal{P}(3\pi,\tau_1&;&3\pi,\tau_2)
    = \frac{|X_{3\pi}|^4}{2}\left[R_S e^{-\Gamma_L \tau_2 -\Gamma_S \tau_1} + R_S
    e^{-\Gamma_L \tau_1-\Gamma_S \tau_2}\right.
    \nonumber\\&&-2|\overline{\eta}_{3\pi}|^2
    \cos(\Delta m (\tau_1-\tau_2))
    e^{-(\overline{\Gamma}+\alpha-\gamma)(\tau_1+\tau_2)}
   \nonumber\\&&
     +|\overline{\eta}_{3\pi}|e^{ -\Gamma_L \tau_1
    -(\overline{\Gamma}+\alpha-\gamma)\tau_2}
    \left(-2|\omega|  \cos(\Delta m \tau_2 -\phi_{3\pi}+\Omega)
    +\frac{4\beta}{|d|} \sin(\Delta m\tau_2-\phi_{3\pi}+\phi_{SW})
     \right)
   \nonumber\\&& \left.
     +|\overline{\eta}_{3\pi}|e^{ -\Gamma_L \tau_2
    -(\overline{\Gamma}+\alpha-\gamma)\tau_1}
    \left(2|\omega|  \cos(\Delta m \tau_1 -\phi_{3\pi}+\Omega)
    +\frac{4\beta}{|d|} \sin(\Delta m\tau_1-\phi_{3\pi}+\phi_{SW})
     \right)\right],\nonumber
     \nonumber \\ &&+\left(|\omega|^2 + \frac{2\gamma}{\Delta \Gamma}
     +\frac{8\beta}{\Delta\Gamma}{\rm Im} [\overline{\eta}_{3\pi}-Y_{3\pi}]\right)
    e^{-\Gamma_L (\tau_1+\tau_2)}
    \nonumber \\&&
 \eea
This completes our analysis. We observe that the inclusion of the
$\epsilon '$ corrections does not affect the functional form of
the decoherence and CPTV effects.


\begin{thebibliography}{99}
\bibitem{footnote1}
The authors are well aware of the 
limitations of a linear master equation, which may not represent
the full quantum-gravity situation. 
The possibility that quantum gravity may be non-linear has already been 
contemplated in earlier papers (see, for instance, \cite{emnnl}), 
but any non-linearity  may not be visible in  a situation like neutral
kaons, where a mean-field  result, parametrised by a (linear) Lindblad
equation, should suffice.
\end{thebibliography}
\end{document}